\documentclass[12pt]{article}
\pdfoutput=1
\usepackage{jheppub}
\usepackage{mathtools,amsmath,amssymb,amsthm,physics,bm,float}
\usepackage[utf8]{inputenc}
\usepackage{graphicx}
\usepackage{enumitem}
\usepackage{hyperref}
\usepackage{pst-node}
\usepackage{tikz-cd} 
\usepackage{soul}
\usepackage[normalem]{ulem}

\begin{document}

\DeclarePairedDelimiter{\ceil}{\lceil}{\rceil}

\theoremstyle{definition}
\newtheorem{definition}{Definition}[section]

\newtheorem{theorem}{Theorem}
\newtheorem{lemma}[theorem]{Lemma}
\newtheorem{proposition}[theorem]{Proposition}
\newtheorem{corollary}[theorem]{Corollary}
\newtheorem{claim}{Claim}

\theoremstyle{remark}
\newtheorem{remark}{Remark}

\newcommand{\be}{\begin{equation}}
\newcommand{\ee}{\end{equation}}
\newcommand{\bi}{\begin{itemize}}
\newcommand{\ei}{\end{itemize}}
\newcommand{\mathbbm}[1]{\text{\usefont{U}{bbm}{m}{n}#1}}
\def\ba#1\ea{\begin{align}#1\end{align}}
\def\bg#1\eg{\begin{gather}#1\end{gather}}
\def\bm#1\em{\begin{multline}#1\end{multline}}
\def\bmd#1\emd{\begin{multlined}#1\end{multlined}}
\setlength{\intextsep}{-1ex} 

\def\XD#1{{\color{magenta}{ [#1]}}}
\def\SAM#1{{\color{red}{ [#1]}}}
\def\WW#1{{\color{blue}{ [#1]}}}
\def\YL#1{{\color{green}{ [#1]}}}

\def\a{\alpha}
\def\b{\beta}
\def\c{\chi}
\def\C{\Chi}
\def\d{\delta}
\def\D{\Delta}
\def\e{\epsilon}
\def\ve{\varepsilon}
\def\g{\gamma}
\def\G{\Gamma}
\def\h{\eta}
\def\k{\kappa}
\def\l{\lambda}
\def\L{\Lambda}
\def\m{\mu}
\def\n{\nu}
\def\p{\phi}
\def\P{\Phi}
\def\vp{\varphi}
\def\q{\theta}
\def\Q{\Theta}
\def\r{\rho}
\def\s{\sigma}
\def\S{\Sigma}
\def\t{\tau}
\def\u{\upsilon}
\def\U{\Upsilon}
\def\w{\omega}
\def\W{\Omega}
\def\x{\xi}
\def\X{\Xi}
\def\y{\psi}
\def\Y{\Psi}
\def\z{\zeta}

\newcommand{\la}{\label}
\newcommand{\ci}{\cite}
\newcommand{\re}{\ref}
\newcommand{\er}{\eqref}
\newcommand{\se}{\section}
\newcommand{\sse}{\subsection}
\newcommand{\ssse}{\subsubsection}
\newcommand{\fr}{\frac}
\newcommand{\na}{\nabla}
\newcommand{\pa}{\partial}
\newcommand{\td}{\widetilde}
\newcommand{\wtd}{\widetilde}
\newcommand{\ph}{\phantom}
\newcommand{\eq}{\equiv}
\newcommand{\wg}{\wedge}
\newcommand{\cd}{\cdots}
\newcommand{\nn}{\nonumber}
\newcommand{\qu}{\quad}
\newcommand{\qqu}{\qquad}
\newcommand{\lt}{\left}
\newcommand{\rt}{\right}
\newcommand{\lra}{\leftrightarrow}
\newcommand{\ol}{\overline}
\newcommand{\ap}{\approx}
\renewcommand{\(}{\left(}
\renewcommand{\)}{\right)}
\renewcommand{\[}{\left[}
\renewcommand{\]}{\right]}
\newcommand{\<}{\langle}
\renewcommand{\>}{\rangle}
\newcommand{\Hc}{\mathcal{H}_{code}}
\newcommand{\HR}{\mathcal{H}_R}
\newcommand{\HRb}{\mathcal{H}_{\ol{R}}}
\newcommand{\lan}{\langle}
\newcommand{\ran}{\rangle}
\newcommand{\Hra}{\mathcal{H}_{W_\alpha}}
\newcommand{\Hrba}{\mathcal{H}_{\ol{W}_\alpha}}

\newcommand{\bH}{{\mathbb H}}
\newcommand{\bR}{{\mathbb R}}
\newcommand{\bZ}{{\mathbb Z}}
\newcommand{\cA}{{\mathcal A}}
\newcommand{\cB}{{\mathcal B}}
\newcommand{\cC}{{\mathcal C}}
\newcommand{\cE}{{\mathcal E}}
\newcommand{\cH}{{\mathcal H}}
\newcommand{\cI}{{\mathcal I}}
\newcommand{\cN}{{\mathcal N}}
\newcommand{\cO}{{\mathcal O}}
\newcommand{\zb}{{\bar z}}

\newcommand{\Area}{\operatorname{Area}}
\newcommand{\ext}{\operatorname*{ext}}
\newcommand{\total}{\text{total}}
\newcommand{\bulk}{\text{bulk}}
\newcommand{\brane}{\text{brane}}
\newcommand{\matter}{\text{matter}}
\newcommand{\Wald}{\text{Wald}}
\newcommand{\anomaly}{\text{anomaly}}
\newcommand{\extrinsic}{\text{extrinsic}}
\newcommand{\gen}{\text{gen}}
\newcommand{\mc}{\text{mc}}

\newcommand{\pGI}{\Pi_\text{GI}}

\newcommand{\T}[3]{{#1^{#2}_{\ph{#2}#3}}}
\newcommand{\Tu}[3]{{#1_{#2}^{\ph{#2}#3}}}
\newcommand{\Tud}[4]{{#1^{\ph{#2}#3}_{#2\ph{#3}#4}}}
\newcommand{\Tdu}[4]{{#1_{\ph{#2}#3}^{#2\ph{#3}#4}}}
\newcommand{\Tdud}[5]{{#1_{#2\ph{#3}#4}^{\ph{#2}#3\ph{#4}#5}}}
\newcommand{\Tudu}[5]{{#1^{#2\ph{#3}#4}_{\ph{#2}#3\ph{#4}#5}}}

\title{Holographic Tensor Networks with Bulk Gauge Symmetries}

\author[a]{Xi Dong,}
\author[a]{Sean McBride,}
\author[a, b]{and Wayne W. Weng}

\affiliation[a]{Department of Physics, University of California, Santa Barbara, CA 93106, USA}
\affiliation[b]{Department of Physics, Cornell University, Ithaca, NY 14853, USA}

\emailAdd{xidong@ucsb.edu}
\emailAdd{seanmcbride@ucsb.edu}
\emailAdd{www62@cornell.edu}

\abstract{Tensor networks are useful toy models for understanding the structure of entanglement in holographic states and reconstruction of bulk operators within the entanglement wedge. They are, however, constrained to only prepare so-called ``fixed-area states'' with flat entanglement spectra, limiting their utility in understanding general features of holographic entanglement. Here, we overcome this limitation by constructing a variant of random tensor networks that enjoys bulk gauge symmetries. Our model includes a gauge theory on a general graph, whose gauge-invariant states are fed into a random tensor network. We show that the model satisfies the quantum-corrected Ryu-Takayanagi formula with a nontrivial area operator living in the center of a gauge-invariant algebra. We also demonstrate nontrivial, $n$-dependent contributions to the R\'enyi entropy and R\'enyi mutual information from this area operator, a feature shared by general holographic states.}

\maketitle

\section{Introduction}

The ultimate goal of the AdS/CFT correspondence is to understand, concretely, the relationship between a bulk gravitational theory and its dual boundary conformal field theory. Holographic duality posits that the partition functions of the two theories are equal and that there exists an isomorphism between the Hilbert space of states of a theory of quantum gravity $\mathcal{H}_{\textrm{bulk}}$ and the Hilbert space of a seemingly unrelated quantum mechanical system $\mathcal{H}_{\textrm{{boundary}}}$. If we were to understand the precise relation between these Hilbert spaces, we would have a tractable handle with which to study quantum gravity, in whatever form it may ultimately arise.

In practice, the UV degrees of freedom in the bulk are not well-understood, so one must often be satisfied with studying a subspace of states given by small fluctuations around a fixed semiclassical saddle. These states span a code subspace of the quantum gravity Hilbert space, and are thus embedded in the larger Hilbert space of the dual boundary theory, in the same way as the logical qubits of a quantum error correcting code (QECC) are embedded in a larger Hilbert space of physical qubits \cite{Almheiri:2014lwa}.

In the last decade, a useful tool for developing intuition about the bulk-to-boundary map has been tensor networks. Tensor networks, specifically  projected entangled pair states (PEPS) and PEPS-inspired tensor networks, originally arose in many-body physics as a generalization of matrix product states, which allowed one to efficiently prepare spin chain states with area law entanglement \cite{Verstraete:2004cf}.

As a toy model for holography, tensor networks found their niche due to the fact that they obey the Ryu-Takayanagi (RT) formula \cite{Ryu:2006bv} and its refinements \cite{Faulkner:2013ana, Swingle:2009bg, Lewkowycz:2013nqa, Dong:2017xht}. In particular, random tensor networks (RTNs) \cite{Hayden:2016cfa} reproduce several desirable properties of a holographic QECC, namely satisfying a quantum-corrected RT formula and the Petz recontruction of local operators \cite{Jia:2020etj}.

We now give a short overview of holographic RTNs and their entanglement properties, as well as their issues. A rank-$k$ tensor can be represented by its components $T_{\mu_1 \cdots \mu_k}$, with $\mu_i = 1, \dots, D_i$ (the bond dimension). We can associate to each leg a $D_i$-dimensional Hilbert space $\mathcal{H}_i$ spanned by an orthonormal basis of states $\{|\mu_i\>, \,\, \mu_i = 1, \cdots, D_i\}$. The tensor $T$ can then be thought of as a state on the tensor product Hilbert space $\bigotimes_{i=1}^k \mathcal{H}_i$:
\be
|T\> = \sum_{\mu_1, \cdots, \mu_k} T_{\mu_1 \cdots \mu_k} |\mu_1\> \otimes \cdots \otimes |\mu_k\>.
\ee
To construct a tensor network, we consider a set of vertices and links which form a network. To each vertex $x$ we associate a state $|T_x\rangle$, such that the collection of all tensors defines a product state $\otimes_x |T_x\>$. Adjacent tensors are those connected by a link; their corresponding legs are contracted by projecting onto a maximally entangled state. For simplicity, we assume that all contracted legs have the same bond dimension $D$.  Denoting the tensor product Hilbert space on the two legs connecting the tensors at vertices $x$ and $y$ as $\mathcal{H}_{xy} \otimes \mathcal{H}_{yx}$, this means that we project onto the state $\ket{xy} = D^{-1/2} \sum_{\mu = 1}^{D} \ket{\mu_{xy}} \otimes \ket{\mu_{yx}}$. Uncontracted legs are called ``dangling'' and come in two types: bulk legs (viewed as input) and boundary legs (viewed as output). We write the boundary state in the following way:\footnote{Here, we have chosen a pure state as the bulk input, but generalizing to mixed states is straightforward.}
\be
\ket{\Psi_\partial} = \left( \bra{\Phi_b} \otimes \bigotimes_{\expval{xy}}  \bra{xy} \right) \left( \bigotimes_x \ket{T_x} \right),
\ee
where we project the bulk input legs onto a bulk state $\ket{\Phi_b}$. In an RTN, we choose $T_x$ to be independent random tensors and take $D$ to be large. We will not go into details on how one computes R\'enyi entropy in the RTN here; the important point is that, for a boundary subregion $R$, one finds the following answer for the R\'enyi entropy $S_n(R)$:
\be
S_n(R) = \abs{\gamma_R} \log D + S_n(\rho_r),
\label{eq:RTNentropy}
\ee
where $\abs{\gamma_R}$ is the number of links cut by the minimal surface $\gamma_R$ homologous to $R$ and $S_n(\rho_r)$ is the R\'enyi entropy of the bulk subregion $r$ bounded by $R \cup \gamma_R$ (we will call $r$ the entanglement wedge). Analytically continuing to $n = 1$ recovers the Faulkner-Lewkowycz-Maldacena (FLM) formula
\be
S_{\textrm{vN}}(R) = \frac{\big\langle \hat{A}  \big\rangle}{4G_N} + S_{\textrm{vN}}(\rho_r),
\label{eq:flm}
\ee
with $\abs{\gamma_R} \log D$ identified with the expectation value of the area operator $\langle \hat{A} \rangle/4G_N$.

\begin{figure}
    \centering
    \includegraphics[width=0.5\textwidth]{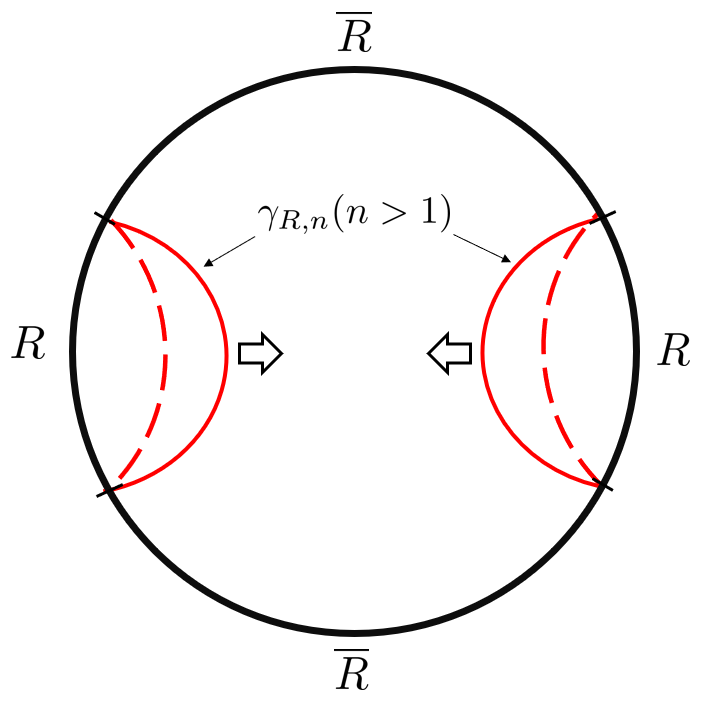}
    \caption{The cosmic branes that arise in computing the R\'enyi entropy for disjoint subregions. These branes have nonzero, $n$-dependent tension, and so would backreact in a realistic holographic system.}
    \label{fig:branes}
\end{figure}

In a state with vanishing bulk R\'enyi entropy (such as a product state), the boundary R\'enyi entropy \er{eq:RTNentropy} is consequently independent of $n$. The RTN thus exhibits a flat entanglement spectrum due to the projection of contracted legs onto maximally mixed states.\footnote{The HaPPY code~\cite{Pastawski:2015qua} also features a flat R\'enyi spectrum for similar reasons.} This differs sharply from what we expect from generic situations in AdS/CFT. For example, the R\'enyi entropy for an interval $R$ of length $\ell$ in the vacuum state of a two-dimensional CFT takes the form
\be
S_n(R) = \frac{c}{6} \left( 1 + \frac{1}{n}\right) \log \left( \frac{\ell}{\epsilon}\right),
\label{eq:CFT}
\ee
which is manifestly $n$-dependent. One possible solution is to instead project contracted legs onto a non-maximally entangled link state \cite{Dong:2021clv,Cheng:2022ori}. By tuning the entanglement spectrum appropriately, this allows one to reproduce the correct single-interval CFT vacuum R\'enyi entropy \er{eq:CFT}, but does not work in more general cases such as that of multiple disjoint intervals. To see this, consider two disjoint intervals $R_1$ and $R_2$ (see Figure \ref{fig:branes}), and for simplicity consider the case where the mutual information between the intervals is small in the sense that the RT surfaces are always in a disconnected phase.
The boundary R\'enyi entropy can be obtained by inserting appropriate cosmic branes into the bulk~\cite{Dong:2016fnf}.
The tension of the cosmic branes is proportional to $1 - 1/n$. In a fully gravitating system, the two cosmic branes homologous to $R_1$, $R_2$ will backreact and affect each other in an $n$-dependent way. This results in a nonzero R\'enyi mutual information between the two intervals that cannot be reproduced in RTNs by simply adding non-maximally entangled links, because they would not allow the minimal surfaces to affect each other. 

From the gravity point of view, the RTN prepares a so-called fixed-area state \cite{Akers:2018fow, Dong:2018seb}, which is an eigenstate of the area operator $\hat{A}$ in \er{eq:flm}. Such eigenstates form a complete basis for semiclassical states prepared via the gravitational path integral, so in principle any semiclassical state can be represented as a superposition over fixed-area basis states $\ket{\alpha}$, where $\alpha$ labels the eigenvalues of the area operator. As the area operator lives on the RT surface dividing the entanglement wedge $r$ and its complement $\overline{r}$, it naturally belongs to the center of the algebra of bulk operators in $r$. This view was espoused in \cite{Harlow:2016vwg}, where it was shown that the FLM formula \er{eq:flm} can be derived from a quantum error correcting code with complementary recovery. In that language, the area operator is a specific element of the center of the bulk von Neumann algebra on $r$. The usual RTN implements a special case of this where the algebra has a trivial center, i.e., the center consists of $c$-numbers only and is therefore isomorphic to $\mathbbm{C}$. In particular, this means that the area operator must be a $c$-number, which, as previously discussed, is incongruous with what one observes in gravitational holography.

The goal of this paper is to construct a model where the algebra on $r$ has a nontrivial center and to identify a nontrivial area operator living in the center.\footnote{Having a nontrivial center does not guarantee that the area operator is not a $c$-number; for example, see \cite{Cao:2023mzo}.} An \textit{ad hoc} way of getting a nontrivial center is to ``stack'' multiple layers of tensor networks by hand to form superpositions of fixed-area states. We will not do this but will instead pursue a more physically motivated approach. In particular, one would like to incorporate something akin to ``edge states'', degrees of freedom which live on the minimal surface, in order to go beyond fixed-area states and produce a nontrivial area operator.\footnote{Initial work in this direction was taken in \cite{Donnelly:2016qqt} by generalizing the HaPPY code.} Our goal in this work is to give a model which provides a physical origin for these edge states. Inspired by similar operators found in gauge theory \cite{Donnelly:2011hn}, we will add a second layer on top of the standard RTN which imposes gauge invariance. This alters the algebra of operators in the bulk, and as we will show, it introduces a nontrivial contribution to the area operator of the following form:
\begin{equation}
    \Delta \widetilde{A} = \bigoplus_{\alpha} \widetilde{P}^\alpha \log d_\alpha,
\end{equation}
where roughly speaking $\alpha$ denotes a superselection sector in the gauge-invariant Hilbert space, $\widetilde{P}^\alpha$ is the projection onto that superselection sector, and $d_\alpha$ is the dimension of $\a$ viewed as an irreducible representation. The important thing to note at the moment is that this operator is not a $c$-number and is therefore nontrivial.

The structure of this paper is as follows. In Section \ref{sec:RTN} we will set up our model -- a two-layer gauged random tensor network -- and introduce the formalism for gauge theory on a graph. In Section \ref{sec:GIalg} we will analyze the Hilbert space of gauge-invariant states and the algebras of gauge-invariant operators for a subregion. In Section \ref{sec:entropy} we will compute entanglement and R\'enyi entropies in both the pre-gauged and gauge-invariant algebras, which we will use to derive the new area operator for our model. We conclude with some discussion and future directions.

\section{The Gauged Random Tensor Network}
\label{sec:RTN}

\begin{figure}
    \centering
    $\vcenter{\hbox{\includegraphics[width=0.35\textwidth]{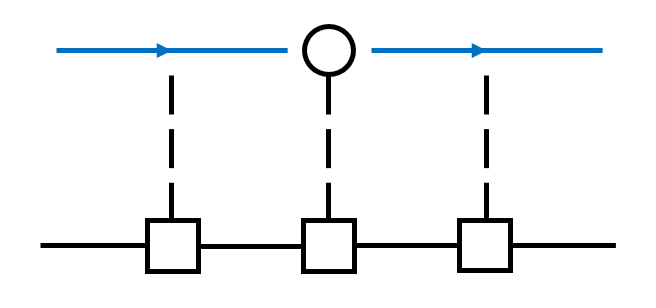}}}$
    $\vcenter{\hbox{\includegraphics[width=0.6\textwidth]{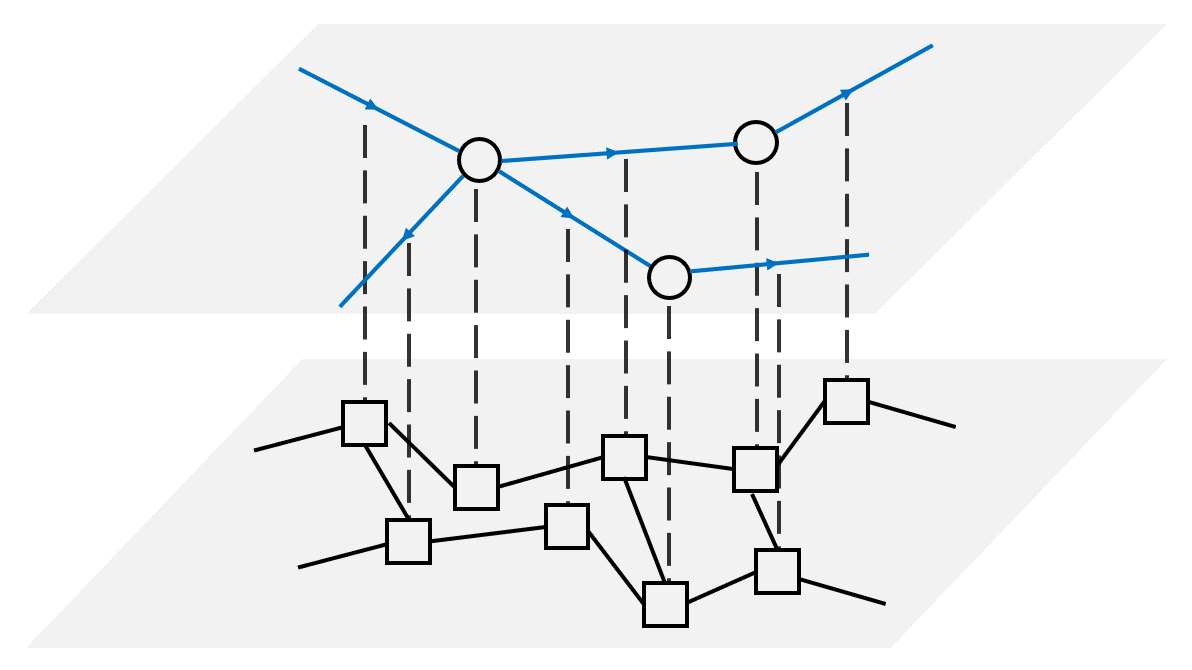}}}$
    \caption{Some examples of our two-layer model, with a gauge theory on a directed graph on the top layer and a random tensor network with dangling boundary legs on the bottom. In these examples, we choose each tensor in the bottom layer to have a bulk input leg which is either a vertex or edge on the graph. The light gray planes in the right example are included for visual clarity.}
    \label{fig:model}
\end{figure}

We now construct our model. It has two layers: a top layer consisting of a gauge theory on a graph, and a bottom layer made of a standard random tensor network.
We illustrate some examples of this two-layer model in Figure \ref{fig:model}.
The top layer produces a gauge-invariant state which is then fed into the bottom layer as input. The final output of the model is the boundary state produced by the bottom RTN. We can then analyze properties of the boundary state (such as its entropy) using the usual techniques for the random tensor network.

This construction has some nice properties. In particular, one might be worried that if the structure of the RTN is altered, Petz reconstruction of local operators might no longer hold. Here we avoid this potential issue by keeping the tensor network the same, but changing the space of states that can be fed into the network. 

Given this construction, we would like to understand what set of gauge-invariant states we will be feeding into the bottom layer. The following is based on a non-dynamical version of the standard Kogut-Susskind construction in lattice gauge theory \cite{Kogut:1974ag}.\footnote{See also related discussion in \cite{Dolev:2021ofc}.} As we do not require our graph to be a lattice, i.e. there is not necessarily a regular tiling, we will refrain from calling our top layer a lattice gauge theory.

Our starting point is an arbitrary directed graph $\Lambda = (V,E)$ consisting of vertices $V = \{ v \}$ and edges $E = \{ e \}$. We require the graph to be directed so we have a well-defined orientation on each edge, though we emphasize that the choice of orientation is arbitrary. We impose no additional conditions on the graph. In particular, the graph could have loops, adjacent vertices could be connected by multiple edges, and the graph does not need to be planar.

We start with a gauge group, which we choose to be a compact Lie group $G$. It does not have to be connected, and in particular, we could consider finite groups such as $\bZ_2$ if we wish. We assign a (pre-gauged) Hilbert space to each vertex and edge of the graph $\Lambda$. The Hilbert space $\mathcal{H}_e$ on each edge $e$ is taken to be $L^2(G)$, the space of square-integrable functions on $G$. A state $\ket{\psi}_e$ in this $\mathcal{H}_e = L^2(G)$ can be written as an integral\footnote{In cases where $G$ is finite, the integral is understood as a sum: $\ket{\psi}_e = \sum_{g \in G} \frac{1}{\abs{G}} \psi(g) \ket{g}_e$, where $|G|$ is the order of $G$.}
over orthonormal basis elements $\ket{g}$ labeled by $g\in G$:
\be
\ket{\psi}_e = \int dg \psi(g) \ket{g}_e,
\ee
where $dg$ is the Haar measure\footnote{The Haar measure on $G$ is invariant under left and right group multiplication ($g \rightarrow g' g$ and $g \rightarrow g g'$) and is normalized such that $\int dg = 1$.} on $G$. For our purposes, it will be useful to work with another orthonormal basis
\be
\ket{\alpha i j}_e, \quad i, j = 1, 2, \cdots, d_\alpha
\la{eq:erepbasis}
\ee
for the same $\mathcal{H}_e = L^2(G)$, where $\a$ labels irreducible representations (irreps) of $G$ and $d_\a$ is the dimension of the representation $\a$. This representation basis is orthonormal:
\be
{}_e \<\alpha i j | \beta k \ell\>_e = \d_{\a\b} \d_{ik} \d_{j\ell},
\ee
and can be written in terms of the previously defined group basis $\ket{g}_e$:
\be
\ket{\alpha i j }_e = \sqrt{d_\alpha} \int dg \, D^\alpha_{ij} (g) \ket{g}_e,
\ee
where $D^\alpha_{ij}(g)$ are elements of a unitary matrix $D^\a(g)$ representing $g$ in $\a$.
This can be viewed as a ``Fourier transform'' between the representation basis and the group basis.

The group action induces a set of unitaries $L_e(g)$ and $R_e(g)$ which act as left and right group multiplications on the group basis:
\be
L_e(g)\ket{h}_e = \ket{gh}_e, \quad R_e(g^{-1})\ket{h}_e = \ket{hg^{-1}}_e .
\la{eq:lrrepbasis}
\ee
In the representation basis, the group unitaries instead act as unitary matrix multiplication on one of the two indices $i$, $j$:
\be
L_e(g)\ket{\alpha ij}_e = \sum_k D^{\ol\alpha}_{ki}(g) \ket{\alpha kj}_e, \quad R_e(g^{-1}) \ket{\alpha ij}_e = \sum_k D^\alpha_{kj}(g) \ket{\alpha ik}_e,
\ee
where $\overline{\a}$ denotes the complex conjugate representation of $\a$ defined by $D^{\ol\alpha}_{ij}(g) = D^{\alpha*}_{ij}(g)$.
Thus there are two copies of $G$ acting on $\mathcal H_e$: $L_e(g)$ gives the group action of the first copy under which $\ket{\alpha ij}_e$ transforms in the representation $\ol\a$, and $R_e(g^{-1})$ gives the action of the second copy under which $\ket{\alpha ij}_e$ transforms in the representation $\a$.
Altogether, $\ket{\alpha ij}_e$ transforms in the external tensor product\footnote{For representations $\alpha_1$, $\alpha_2$ of $G$, their external tensor product $\alpha_1 \boxtimes \alpha_2$ is a representation of $G \times G$ with an underlying vector space $\cH^{\alpha_1} \otimes \cH^{\alpha_2}$, where $\cH^{\alpha_1}$ transforms under the first $G$ in the $\alpha_1$ representation and $\cH^{\alpha_2}$ transforms under the second $G$ in the $\alpha_2$ representation. Note that this is different from the (usual) tensor product $\alpha_1 \otimes \alpha_2$ which is a representation of $G$ (not $G \times G$), with an underlying vector space $\cH^{\alpha_1} \otimes \cH^{\alpha_2}$ where $\cH^{\alpha_1}$ and $\cH^{\alpha_2}$ transform under the same $G$.} representation $\ol\a \boxtimes \a$ of $G \times G$.
Using this, we decompose $\mathcal{H}_e$ as
\be
\mathcal{H}_e \cong \bigoplus_\alpha \mathcal{H}^{\ol\alpha} \otimes \mathcal{H}^{\alpha} \cong \bigoplus_\alpha \left( \mathcal{H}^\alpha \right)^{\oplus d_\alpha},
\label{eq:l2giso}
\ee
where the sum runs over all irreducible representations $\alpha$ of $G$ and $\mathcal H^\a$ is a Hilbert space of dimension $d_\a$ transforming in the $\a$ representation.
It will be convenient to use the representation basis for the remainder of the paper.

Now we turn to the (pre-gauged) Hilbert space $\cH_v$ on a vertex $v$.
In general, $\cH_v$ may be chosen quite arbitrarily (corresponding to specifying any number of matter degrees of freedom including the case of no matter), but it needs to furnish some representation under the group action of $G$. This representation could be reducible or trivial, but it can always be decomposed into a direct sum of irreducible representations of $G$.
Using this, we may decompose a general $\cH_v$ as
\be
\mathcal{H}_v = \bigoplus_{\alpha} \left( \mathcal{H}_v^\alpha \right)^{\oplus n_\alpha}.
\label{eq:vertexHilbert}
\ee
Here the sum again runs over all distinct irreducible representations $\alpha$ of $G$ and $n_\alpha$ is the multiplicity of the representation $\a$ in $\cH_v$. Note that $n_\a$ could be any nonnegative integer, and in particular, it could be zero (representing the absence of a given representation $\a$ in $\cH_v$). Thus, the simplest choice of $\cH_v$ is a trivial Hilbert space with no matter (corresponding to $n_\a=0$ for all $\a$), but in the discussion below we will consider the general case \er{eq:vertexHilbert} with arbitrary $n_\a$. Furthermore, we will allow $\cH_v$ to vary from one vertex to another.
An orthonormal basis of states for the Hilbert space $\cH_v$ can be written as
\be
\ket{\alpha i j}_v, \quad i = 1, \cdots, n_\alpha, \quad j = 1, \cdots, d_\alpha,
\la{eq:vbasis}
\ee
where the first index $i$ runs over the multiplicity $n_\a$ and the second runs over the dimension $d_\a$.
The group action of $G$ on $\cH_v$ is given by unitary operators $U_v(g)$, which act on the $\ket{\alpha i j}_v$ basis as
\be
U_v(g) \ket{\alpha i j}_v = \sum_k D^\alpha_{kj}(g) \ket{\alpha i k}_v.
\ee
Note that $U_v(g)$ only acts on the second index $j$ and is analogous to the action of $R_e(g^{-1})$ in \er{eq:lrrepbasis}. Thus, we find an important distinction between the vertex Hilbert space $\cH_v$ and the edge Hilbert space $\cH_e$. To see this, first note that the two Hilbert spaces share some similarities. In particular, $\cH_e$ is a direct sum of irreducible representations $\a$ with multiplicity $d_\a$ as shown on the right-hand side of \eqref{eq:l2giso}, and this is the analogue of \er{eq:vertexHilbert} for $\cH_v$. The representation basis \er{eq:erepbasis} of $\cH_e$ is similar to the basis \er{eq:vbasis} of $\cH_v$. However, the difference is that an edge has the additional structure of allowing another group action $L_e(g)$ that acts on the first index $i$ of $\ket{\a ij}_e$, whereas at a vertex the first index $i$ of $\ket{\a ij}_v$ is a multiplicity index that does not admit a natural group action.

The pre-gauged Hilbert space for the entire graph is then
\be
\mathcal{H} = \left( \bigotimes_{v \in V} \mathcal{H}_v \right) \otimes  \left( \bigotimes_{e \in E} \mathcal{H}_e \right).
\ee
We refer to the algebra of all bounded operators on $\cH$ as $\mathcal{A} = \mathcal{B} \left( \mathcal{H} \right)$. As $\cH$ completely factorizes over the vertices and edges, so too does the algebra of operators
\begin{equation}
    \mathcal{A} = \left(\bigotimes_{v \in V} \mathcal{A}_v\right) \otimes \left( \bigotimes_{e \in E} \mathcal{A}_e \right).
\end{equation}
Using the representation basis \eqref{eq:erepbasis} of $\cH_e$, $\mathcal{A}_e$ can be written as
\be
\mathcal{A}_e = \textrm{span} \{ \ket{\alpha i j}_e \bra{\beta k \ell}\},
\ee
where the indices $i,j,k,\ell$ run over the irrep dimension.
Similarly, using \er{eq:vbasis} we write
\begin{equation}
\mathcal{A}_v = \textrm{span} \{ \ket{\alpha i j}_v \bra{\beta k \ell} \},
\end{equation}
where $i,k$ run over the irrep multiplicity and $j,\ell$ run over the irrep dimension.

For each vertex $v$, we now define a gauge transformation $A_v(g)$ as the following unitary operator acting on $v$ and all its associated edges:
\be
A_v(g) \equiv U_v(g) \prod_{e \in E^{-}(v)} L_e(g) \prod_{e \in E^{+}(v)} R_e(g^{-1}),
\la{eq:avdef}
\ee
where $E^{-}(v)$ consists of edges associated to $v$ oriented away from the vertex and $E^{+}(v)$ consists of edges oriented into the vertex. Physical states are defined to be those invariant under gauge transformations $A_v(g)$ for all $g$ and $v$. The easiest way of generating a gauge-invariant state is to average over all gauge transformations acting on a state in $\cH$. The operator that implements this averaging on a vertex $v$ is the following projector:
\be
\Pi_v = \int dg A_v(g) .
\ee
$\Pi_v$ obeys the usual properties of a projector such that $\Pi_v^2 = \Pi_v$ and $\Pi_v = \Pi_v^\dagger$. The gauge-invariant projector on the entire graph is simply the product of individual projectors on all vertices:
\be
\pGI = \prod_{v \in V} \Pi_v.
\ee
It is easy to verify that $[A_v(g), A_{v'}(g')]=0$ for all $v$, $v'$, $g$, $g'$, and therefore $[\Pi_v, \Pi_{v'}]=0$.

Throughout the paper, we will denote fully gauge-invariant spaces, states, and operators with a tilde; for instance, the gauge-invariant states $\ket{\widetilde{\psi}}$ are elements of $\widetilde{\mathcal{H}}$ defined via
\be
\widetilde{\mathcal{H}} \equiv \pGI \mathcal{H}.
\ee
The gauge-invariant algebra $\widetilde{\mathcal{A}}$ is defined as the space of bounded operators on $\widetilde{\mathcal{H}}$. $\widetilde{\mathcal{A}}$ can alternatively be represented by conjugation of the pre-gauged algebra $\cA$ with the projector $\pGI$:
\be
\widetilde{\mathcal{A}} = \pGI \mathcal{A} \pGI.
\label{eq:gaugedalg}
\ee
We should comment on the interpretation of the operators in this gauge-invariant algebra. Every operator $\widetilde{\mathcal{O}} \in \widetilde{\mathcal{A}}$ can be extended to a pre-gauged operator $\mathcal{O} \in \mathcal{A}$ which acts identically on gauge-invariant states. There is generally more than one extension to $\mathcal{A}$, and to choose a unique extension one must specify the action of the pre-gauged operator on the orthogonal complement of $\widetilde{\mathcal{H}}$. We make the natural choice that the extension $\cO$ should annihilate the orthogonal complement. Moreover, for notational simplicity, we identify every $\widetilde{\mathcal{O}} \in \widetilde{\mathcal{A}}$ with its natural extension $\cO \in \cA$ (which annihilates the orthogonal complement), as we have done in \er{eq:gaugedalg}. The reason for this natural extension will become clearer in later sections. 

We now feed any gauge-invariant state $\ket{\widetilde{\psi}}$ as the bulk input into the RTN on the the bottom layer, in a manner illustrated by Figure \ref{fig:model}.
In particular, the bulk dangling legs of the RTN should match and connect to the edges and vertices of the graph $G$ on the top layer, for $\ket{\widetilde{\psi}}$ lives on these edges and vertices. In other words, each edge or vertex of $G$ is fed into a bulk dangling leg of the RTN.\footnote{In principle, the RTN could also take any pre-gauged state as the bulk input, but we choose to feed only gauge-invariant states because as we will see, this restriction leads to a nontrivial area operator.}

In order to utilize the full machinery of the original RTN, we would like the Hilbert spaces associated with the tensors on the bottom layer to be finite-dimensional (as is the case for the original RTN). When $G$ is an infinite group, $\cH_e = L^2(G)$ is infinite-dimensional and there are an infinite number of irreducible representations to sum over, so in order to avoid a tensor in the bottom layer having an infinite-dimensional leg, we impose a cutoff on our edge and vertex Hilbert spaces. This can take the form of, e.g., a cutoff in the sums in \eqref{eq:l2giso} and \eqref{eq:vertexHilbert}. Therefore, we are only feeding in states that live in a finite-dimensional subspace of $\widetilde{\mathcal{H}}$. This does not affect the discussion in the next section of the gauge-invariant algebra; the cutoff is only relevant when we compute entanglement measures in Section \ref{sec:entropy}.

\section{Deriving the Gauge-Invariant Algebra}
\label{sec:GIalg}

Now that we have defined our gauge-invariant states, we would like to understand the structure of the algebra of gauge-invariant operators. Our overarching goal is to write down the gauge-invariant subalgebra for a subregion $r$ of the top layer which we will later use to derive an FLM formula for the gauged RTN.

\subsection{The structure of the gauge-invariant Hilbert space}

We now study the decomposition of $\wtd\cH$ when our graph $\Lambda$ is divided into a subregion and its complement. We define a subregion $r$ of $\Lambda$ to be an arbitrary subset of vertices and edges (without further restrictions). We call the complement subregion $\overline{r}$. 

In order to work out a useful basis for gauge-invariant states, it is convenient to divide the set $V$ of all vertices into three types: those strictly in $r$ (meaning that the vertex and its associated edges are all in $r$), those strictly in $\overline{r}$, and vertices ``on the cut'' (meaning that the vertex and its associated edges are partly in $r$ and partly in $\overline{r}$). We call these sets $V_r$, $V_{\overline{r}}$, and $V_c \equiv V/\left( V_r \cup V_{\overline{r}}\right)$, respectively. Consequently, the gauge-invariant projector can be decomposed in the following way:
\be
\pGI = \Pi_{V_r} \Pi_{V_c} \Pi_{V_{\overline{r}}},
\label{eq:pidecomp}
\ee
where $\Pi_{V_i}$ is defined as the product of individual projections $\Pi_v$ over all vertices $v \in V_i$, for $i=r,c,\ol{r}$. First, let us discuss a partial gauging of the pre-gauged Hilbert space. Using the tensor decomposition of $\mathcal{H}=\mathcal{H}_r \otimes \mathcal{H}_{\overline{r}}$, we can write $\widetilde{\mathcal{H}} = \pGI \cH$ as
\ba
\widetilde{\mathcal{H}} &= \Pi_{V_r} \Pi_{V_c} \Pi_{V_{\overline{r}}} \left( \mathcal{H}_r \otimes \mathcal{H}_{\overline{r}}\right) \nonumber \\
&= \Pi_{V_c} \lt(\left( \Pi_{V_r} \mathcal{H}_r\right) \otimes  \left( \Pi_{V_{\overline{r}}} \mathcal{H}_{\overline{r}} \right)\rt).
\ea
We define the two terms in the parentheses as
\be
\hat{\mathcal{H}}_r \equiv \Pi_{V_r} \mathcal{H}_r, \quad \hat{\mathcal{H}}_{\overline{r}} \equiv \Pi_{V_{\overline{r}}} \mathcal{H}_{\overline{r}}.
\ee
These are ``partially gauged'' Hilbert spaces, in the sense that states in $\hat{\mathcal{H}}_r$ ($\hat{\mathcal{H}}_{\overline{r}}$) are invariant under gauge transformations associated to vertices in $V_r$ ($V_{\ol r}$), but not so under gauge transformations on the cut.
We denote the partially gauged Hilbert space on the full graph as
\be
\hat{\mathcal{H}} = \hat{\mathcal{H}}_r \otimes \hat{\mathcal{H}}_{\overline{r}}.
\ee
As $\hat{\mathcal{H}}$ tensor factorizes, the algebra of operators on $\hat{\mathcal{H}}$ also factorizes as
\be
\hat{\mathcal{A}} = \hat{\mathcal{A}}_r \otimes \hat{\mathcal{A}}_{\overline{r}}.
\ee

Now that we have a partially gauged Hilbert space $\hat{\cH}$, it remains to impose gauge invariance ``on the cut'' and obtain the fully gauged Hilbert space $\wtd\cH = \Pi_{V_c} \hat{\cH}$.
The gauge transformation \er{eq:avdef} associated to each vertex $v_i \in V_c$ can be decomposed into unitary operators in $r$ and $\overline{r}$:
    \be
    A_{v_i} (g_i) = A_{v_i, r} (g_i) A_{v_i, \overline{r}} (g_i).
    \ee
Let $n\eq \abs{V_c}$ be the number of vertices on the cut.
The gauge-invariant projector on the cut $\Pi_{V_c}$ acts by integrating over the gauge transformations associated to the $n$ vertices in $V_c$:
\ba
\Pi_{V_c} &= \int dg_1 \cdots dg_n A_{v_1}(g_1) \cdots A_{v_n}(g_n) \nonumber \\
&= \int dg_1 \cdots dg_n A_{v_1,r}(g_1) \cdots A_{v_n,r}(g_n) A_{v_1,\overline{r}} (g_1) \cdots A_{v_n,\overline{r}} (g_n) \nonumber \\
&\equiv \int dg A_r(g) A_{\overline{r}} (g),
\ea
where we have defined $A_r(g) = \prod_{i = 1}^n A_{v_i, r}(g_i)$ (and similarly for $A_{\overline{r}}(g)$), $g = (g_1, \cdots, g_n)$ is a element of $G^n$ (the direct product of $n$ copies of $G$ on the cut), and $dg$ is the Haar measure on $G^n$. Thus $A_r(g)$ is a $G^n$ action on $\hat{\mathcal{H}}_r$, and $\hat{\mathcal{H}}_r$ can be decomposed into irreps of $G^n$. We decompose $\hat{\mathcal{H}_r}$ into the following way:
\be
\hat{\mathcal{H}}_r \cong \bigoplus_{\alpha,  i} \hat{\mathcal{H}}^{\alpha i}_r,
\label{eq:hhatdecomp}
\ee
where $\alpha$ as an irreducible representation of $G^n$ can also be thought of as the external tensor product of $n$ irreps of $G$, i.e., $\alpha$ denotes the external tensor product $\alpha_1 \boxtimes \alpha_2 \boxtimes \cdots \boxtimes \alpha_n$. Thus, we will sometimes write $\a$ as a tuple of $G$ irreps $(\alpha_1, \alpha_2, \cdots, \alpha_n)$.  The index $i = 1, \cdots, n_\alpha$ denotes the multiplicity of the $\a$ irrep. The sum ranges over all $G^n$ irreps but some irreps may appear with zero multiplicity, as in the single vertex Hilbert space \eqref{eq:vertexHilbert}.

From the decomposition \eqref{eq:hhatdecomp}, we write an orthonormal basis for $\hat{\mathcal{H}}_r$ as $\{ \ket{\alpha i k}_r \} $, where again the first index $i = 1, \cdots, n_\alpha$ runs over the irrep multiplicity and the second index $k = 1, \cdots, d_\alpha$ labels an orthonormal basis for each $\hat{\mathcal{H}}_r^{\alpha i}$. Similarly, we write an orthonormal basis for $\hat{\mathcal{H}}_{\ol r}$ as $\{\ket{\ol{\b} j \ell}_{\ol r}\}$, where $j=1,\cd,{\ol n}_{\ol\b}$, and ${\ol n}_{\ol\b}$ is the multiplicity of the $\ol \b$ irrep on $\ol r$.
Explicitly, $A_r(g)$ ($A_{\overline{r}}(g)$) acts on the basis states of $\hat{\mathcal{H}}_r$ ($\hat{\mathcal{H}}_{\overline{r}}$) via
\ba
A_r(g) \ket{\alpha i k}_r &= \sum_{k'} D^{\alpha}_{k' k} (g) \ket{\alpha i k'}_r, \nonumber \\
A_{\overline{r}}(g) \ket{\overline{\beta} j \ell}_{\overline{r}} &= \sum_{\ell'} D^{\overline{\beta}}_{\ell' \ell} (g) \ket{\overline{\beta} j \ell'}_{\overline{r}}.
\ea
Combining the basis for $\hat{\mathcal{H}}_r$ and for $\hat{\mathcal{H}}_{\ol r}$, we write an orthonormal basis for $\hat{\mathcal{H}}$ as $\{ \ket{\alpha i k}_r \ket{\overline{\beta} j \ell}_{\overline{r}} \}$. It is worth noting that the multiplicities ${\ol n}_{\ol\a}$ on $\ol r$ are generally independent from the multiplicities $n_{\ol \a}$ on $r$; in particular, ${\ol n}_{\ol\a}$ could vanish while $n_{\ol \a}$ is nonzero, and vice versa.

In a sense, we have done as much gauging as we can while keeping the factorization of the Hilbert space between $r$ and $\overline{r}$. $\hat{\mathcal{H}}$ is similar to what is often called the extended Hilbert space \cite{Donnelly:2011hn, Donnelly:2014gva, Aoki:2015bsa, Ghosh:2015iwa}, which is a choice of Hilbert space into which one can embed gauge-invariant states such that the extended Hilbert space factorizes across the cut. Here we arrive at a similar prescription by restricting from a larger Hilbert space $\cH$.

Now we will write a basis of states for the fully gauge-invariant Hilbert space $\widetilde{\mathcal{H}}$. 

\begin{lemma}
The fully gauge-invariant Hilbert space $\wtd\cH = \Pi_{V_c} \left( \hat{\mathcal{H}}_r \otimes \hat{\mathcal{H}}_{\overline{r}}\right)$ is given by
\begin{equation}
    \widetilde{\mathcal{H}} = \left\{ \sum_{\alpha i j k} \widetilde{\psi}_{\alpha i j} \ket{\alpha i k}_r \ket{\overline{\alpha} j k}_{\overline{r}} :  \widetilde{\psi}_{\alpha i j} \in \mathbbm{C} \right\}.
    \label{eq:span1}
\end{equation}
\begin{proof}
    Since we already have a basis for the partially gauged Hilbert space, it suffices to demonstrate the action of $\Pi_{V_c}$ on these basis states, which is given by
\be
\Pi_{V_c} \ket{\alpha i k}_r \ket{\overline{\beta} j \ell}_{\overline{r}} = \sum_{k' \ell'} \int dg  D^{\alpha}_{ k' k}(g) D^{\overline{\beta}}_{\ell' \ell}(g) \ket{\alpha i k'}_r \ket{\overline{\beta} j \ell'}_{\overline{r}}.
\ee
We recall the Schur orthogonality relation for compact groups:
\be
\int dg D^\alpha_{k' k}(g) D^{\overline{\beta}}_{\ell' \ell}(g) = \frac{\delta_{\alpha \beta} \delta_{k' \ell'} \delta_{k \ell}}{d_\alpha},
\ee
so that the fully gauge-invariant basis states are
\be
\Pi_{V_c} \ket{\alpha i k}_r \ket{\overline{\beta} j \ell}_{\overline{r}} = \frac{1}{d_\alpha} \delta_{\alpha \beta} \delta_{k \ell} \sum_{k' \ell'} \delta_{k' \ell'} \ket{\alpha i k'}_r \ket{\overline{\beta} j \ell'}_{\overline{r}} = \frac{1}{d_\alpha} \delta_{\alpha \beta} \delta_{k \ell} \sum_{k'} \ket{\alpha i k'}_r \ket{\overline{\beta} j k'}_{\overline{r}}.
\la{eq:Vcaction}
\ee
Choosing $\alpha = \beta$ and $k = \ell$ gives the desired form \eqref{eq:span1}.
\end{proof}

\end{lemma}
\begin{remark}
\er{eq:span1} immediately implies a natural Hilbert space isomorphism
\be
\widetilde{\mathcal{H}} \cong \bigoplus_{\alpha} \widetilde{\mathcal{H}}_r^\alpha \otimes \widetilde{\mathcal{H}}_{\overline{r}}^{\overline{\alpha}}.
\label{eq:hilbertISO}
\ee
Here $\widetilde{\mathcal{H}}_r^\alpha$ denotes a Hilbert space of dimension $n_\a$ with orthonormal basis states $\ket{\alpha i}_r$ transforming in the $\a$ representation of $G^n$, and $\widetilde{\mathcal{H}}_{\overline{r}}^{\overline{\alpha}}$ similarly denotes a Hilbert space of dimension $\overline{n}_{\overline{\alpha}}$ with orthonormal basis states $\ket{\overline{\alpha} j}_{\overline{r}}$ transforming in the $\ol\a$ representation. Note that although irrep labels such as $\a$ appear in the basis states, they are fixed within each Hilbert space $\widetilde{\mathcal{H}}_r^\alpha$ or $\widetilde{\mathcal{H}}_{\overline{r}}^{\overline{\alpha}}$. 

More explicitly, the natural isomorphism \er{eq:hilbertISO} maps an arbitrary state of \er{eq:span1} in the following way:
\be
\ket{\widetilde{\psi}} = \sum_{\alpha i j k} \widetilde{\psi}_{\alpha i j} \ket{\alpha i k}_r \ket{\overline{\alpha} j k}_{\overline{r}} \qu\rightarrow\qu \sum_{\alpha i j} \sqrt{d_\alpha} \widetilde{\psi}_{\alpha i j} \ket{\alpha i}_r \ket{\overline{\alpha} j}_{\overline{r}}.
\label{eq:stateISO}
\ee
The $\sqrt{d_\alpha}$ is a crucial factor which ensures that the isomorphism preserves the inner product.
\end{remark}

Given this decomposition, our next goal will be to define an algebra of gauge-invariant operators on $r$, which we will call $\widetilde{\mathcal{A}}_r$. Given the lack of factorization of $\widetilde{\mathcal{H}}$ as indicated by \eqref{eq:hilbertISO}, we cannot easily write $\widetilde{\mathcal{A}}_r$ as $\mathcal{B} ( \widetilde{\mathcal{H}}_r )$ for some putative Hilbert space $\widetilde{\mathcal{H}}_r$. Rather, we will use the known algebra of operators on $\mathcal{H}_r$ and $\hat{\mathcal{H}}_r$ to define $\widetilde{\mathcal{A}}_r$.

\subsection{The gauge-invariant subregion algebra}

It is tempting to define the algebra of gauge-invariant operators in a subregion $r$ via restriction of the pre-gauged algebra in that region
\begin{equation}
    \widetilde{\mathcal{A}}_r = \pGI \mathcal{A}_r \pGI,
    \label{eq:gaugedsubalg}
\end{equation}
similar to \eqref{eq:gaugedalg}. There is a second possible description of the gauge-invariant algebra, which is that $\widetilde{\mathcal{A}_r}$ consists of the set of operators $\{ \widetilde{\mathcal{O}}_r = \mathcal{O}_r \pGI \}$ for all operators $\mathcal{O}_r \in \mathcal{A}_r$ which commute with the gauge-invariant projector: $[ \mathcal{O}_r, \pGI ] = 0$. We will call this algebra $\widetilde{\mathcal{A}}_r^{(1)}$, and the algebra \eqref{eq:gaugedsubalg} defined by conjugation by the gauge-invariant projector $\widetilde{\mathcal{A}}_r^{(2)}$. At first blush it is only obvious that $\widetilde{\mathcal{A}}_r^{(1)}$ is a subset of $\widetilde{\mathcal{A}}_r^{(2)}$, as
\be
\mathcal{O}_r \pGI = \mathcal{O}_r \pGI^2 = \pGI \mathcal{O}_r \pGI \Rightarrow \widetilde{\mathcal{A}}_r^{(1)} \subseteq \widetilde{\mathcal{A}}_r^{(2)},
\ee
but it is not obvious the two definitions are equivalent. Here we aim to show that.

\begin{lemma}
$\widetilde{\mathcal{A}}_r^{(1)} = \widetilde{\mathcal{A}}_r^{(2)}$.
    \begin{proof}
        
We again use the group action on the cut $A(g) = A_r(g)A_{\overline{r}}(g)$ and the gauge-invariant projector on the cut $\Pi_{V_c}$ which integrates over the group action:
\be
\Pi_{V_c} = \int dg A(g).
\ee
We define an element of $\widetilde{\mathcal{A}}_r^{(2)}$ by acting on an arbitrary pre-gauged operator $\mathcal{O}_r \in \cA_r$ via
\ba
\pGI \mathcal{O}_r \pGI &= \Pi_{V_{\overline{r}}} \left( \Pi_{V_c} \Pi_{V_r}
\mathcal{O}_r \Pi_{V_r} \Pi_{V_c} \right) \Pi_{V_{\overline{r}}}  \nonumber \\ 
&= \left( \Pi_{V_c} \Pi_{V_r} \mathcal{O}_r \Pi_{V_r} \Pi_{V_c} \right) \Pi_{V_{\overline{r}}} \nonumber \\
&\equiv \left( \Pi_{V_c}\hat{\mathcal{O}}_r  \Pi_{V_c} \right) \Pi_{V_{\overline{r}}}
\ea
where $\hat{\mathcal{O}}_r \eq \Pi_{V_r} \mathcal{O}_r \Pi_{V_r} \in \hat{\mathcal{A}}_r$ is an operator on the partially gauged Hilbert space $\hat{\mathcal{H}}_r$. Conjugation via the gauge-invariant projector on the cut yields
\be
\Pi_{V_c} \hat{\mathcal{O}}_r \Pi_{V_c} = \int dg dg' A(g)  \hat{\mathcal{O}}_r A(g').
\ee
Using the right-invariance of the Haar measure, we can shift $g \rightarrow g (g')^{-1}$ to obtain
\ba
\Pi_{V_c} \hat{\mathcal{O}}_r \Pi_{V_c} &= \int dg A(g) \int dg' A((g')^{-1})  \hat{\mathcal{O}}_r A(g') \\
&= \Pi_{V_c} \int dg' A((g')^{-1}) \hat{\mathcal{O}}_r A(g') \equiv \Pi_{V_c} \hat{\mathcal{O}}_r',
\ea
where $\hat{\mathcal{O}}_r'$ is defined by the integral over $g'$. We could equivalently send $g' \rightarrow g^{-1} g' $ to obtain
\ba
\Pi_{V_c} \hat{\mathcal{O}}_r \Pi_{V_c} &= \int dg A(g)  \hat{\mathcal{O}}_r A(g^{-1}) \Pi_{V_c} \nonumber \\
&= \int dg A(g^{-1}) \hat{\mathcal{O}}_r A(g) \Pi_{V_c} = \hat{\mathcal{O}}_r' \Pi_{V_c} 
\ea
where we use the fact that the Haar measure is invariant under inversion $dg \rightarrow d(g^{-1})$. This shows $\hat{\mathcal{O}}_r' \Pi_{V_c} = \Pi_{V_c} \hat{\mathcal{O}}_r'$, so $\hat{\mathcal{O}}_r'$ commutes with the gauge-invariant projector on the cut. By construction, $\hat{\mathcal{O}}_r'$ also commutes with $\Pi_{V_r}$ and $\Pi_{V_{\ol r}}$, so it commutes with $\pGI$.

Now we show that $\hat{\mathcal{O}}_r'$ is an element of $\mathcal{A}_r$, which is not obvious as $A(g)$ on the cut acts on both $r$ and $\ol r$. However, we can write
\ba
\hat{\mathcal{O}}_r'&=\int dg A(g^{-1})  \hat{\mathcal{O}}_r A(g) = \int dg A_r(g^{-1}) A_{\overline{r}}(g^{-1})   \hat{\mathcal{O}}_r A_{\overline{r}} (g) A_r(g)\nonumber \\
&=  \int dg A_r(g^{-1})  \hat{\mathcal{O}}_r A_r(g),
\ea
as $\hat{\mathcal{O}}_r$ commutes with operators in $\overline{r}$. Thus $\hat{\mathcal{O}}_r'$ is in $\mathcal{A}_r$.

Combining the above, we can write any element of $\widetilde{\mathcal{A}}_r^{(2)}$ as
\be
\pGI \mathcal{O}_r \pGI = \hat{\mathcal{O}}_r' \Pi_{V_c} \Pi_{V_{\overline{r}}} = \hat{\mathcal{O}}_r' \Pi_{V_r} \Pi_{V_c} \Pi_{V_{\overline{r}}} =  \hat{\mathcal{O}}_r' \pGI,
\ee
which belong to $\widetilde{\mathcal{A}}_r^{(1)}$ as $\hat{\mathcal{O}}_r'$ is an operator in $\mathcal{A}_r$ that commutes with $\pGI$.
Therefore, $\widetilde{\mathcal{A}}_r^{(2)} \subseteq \widetilde{\mathcal{A}}_r^{(1)}$. Morevoer, as we argued earlier, we have $\widetilde{\mathcal{A}}_r^{(1)} \subseteq \widetilde{\mathcal{A}}_r^{(2)}$. Thus, we have shown $\widetilde{\mathcal{A}}_r^{(1)} = \widetilde{\mathcal{A}}_r^{(2)}$.
\end{proof}
\begin{remark}
It will be more convenient to use $\widetilde{\mathcal{A}}_r^{(1)}$ as our definition of $\widetilde{\mathcal{A}}_r$ in later discussions. We now rewrite it by introducing the following notation. For the rest of the paper, we will denote the subset of operators in an algebra that commute with the gauge-invariant projector with a superscript $\Pi$; for example, the algebra $\mathcal{A}^\Pi$ is defined by
\be
\mathcal{A}^\Pi \equiv \{ \mathcal{O} \in \mathcal{A} : [\mathcal{O}, \pGI] = 0 \}.
\ee
It is clear that this subset is itself a von Neumann algebra, as it contains the identity, which necessarily commutes with any projector, and is closed under addition, multiplication, and involution\footnote{Closure of $\mathcal{A}^\Pi$ under addition and multiplication is obvious, and closure under involution follows from the projector being Hermitian.}.
Similarly, we define the subalgebra $\mathcal{A}_r^\Pi$ as
\be
\mathcal{A}_r^\Pi = \{ \mathcal{O}_r \in \mathcal{A}_r : [ \mathcal{O}_r, \pGI] = 0\} ,
\ee
and define $\hat{\mathcal{A}}_r^\Pi$ as
\be
\hat{\mathcal{A}}_r^\Pi = \{ \hat{\mathcal{O}}_r \in \hat{\mathcal{A}}_r : [ \hat{\mathcal{O}}_r, \pGI] = 0\} .
\ee
So far, we have shown
\be
\widetilde{\mathcal{A}}_r = \mathcal{A}_r^\Pi \pGI.
\ee
    \end{remark}
\end{lemma}

\begin{lemma}
    $\mathcal{A}_r^\Pi \pGI = \hat{\mathcal{A}}_r^\Pi \pGI$.
    \begin{proof}
        It is clear that $\hat{\mathcal{A}}_r^\Pi \pGI \subseteq \mathcal{A}_r^\Pi \pGI$, so we only need to show the opposite inclusion. Consider any operator $\mathcal{O}_r \in \mathcal{A}_r^\Pi$. As this operator commutes with $\pGI$, we can use the decomposition of the gauge-invariant projector to write $\mathcal{O}_r \pGI$ as
\be
\mathcal{O}_r \pGI = \pGI \mathcal{O}_r \pGI = \pGI \left( \Pi_{V_r} \mathcal{O}_r \Pi_{V_r} \right) \pGI = \left( \Pi_{V_r} \mathcal{O}_r \Pi_{V_r} \right) \pGI.
\ee
Note that $\Pi_{V_r} \mathcal{O}_r \Pi_{V_r}$ is an operator on $\hat{\mathcal{H}}_r$ that commutes with $\pGI$, so it belongs to $\hat{\mathcal{A}}_r^\Pi$. Thus, every element of $\mathcal{A}_r^\Pi \pGI$ is an element of $\hat{\mathcal{A}}_r^\Pi \pGI$. This shows the inclusion $\mathcal{A}_r^\Pi \pGI \subseteq \hat{\mathcal{A}}_r^\Pi \pGI$, from which we conclude $\mathcal{A}_r^\Pi \pGI = \hat{\mathcal{A}}_r^\Pi \pGI$.
    \end{proof}
    \begin{corollary}
        $\widetilde{\mathcal{A}}_r = \hat{\mathcal{A}}_r^\Pi \pGI$.
    \end{corollary}
\end{lemma}

Using the corollary above, we will now construct a generic operator in $\widetilde{\mathcal{A}}_r$.

\begin{lemma}
$\widetilde{\mathcal{A}}_r$ can be written in the following two forms:
    \ba
\widetilde{\mathcal{A}}_r &= \lt\{ \hat{\mathcal{O}}_r \pGI : \hat{\mathcal{O}}_r = \sum_{\alpha i j k} \hat{\mathcal{O}}_{\alpha i j} \ket{\alpha i k}_r \bra{\alpha j k} \otimes \mathbbm{1}_{\overline{r}},\, \hat{\mathcal{O}}_{\alpha i j} \in \mathbbm{C} \rt\} \label{eq:gaugedmid} \\
&= \lt\{ \widetilde{\mathcal{O}}_r = \sum_{\alpha i i' j k \ell} \widetilde{\mathcal{O}}_{\alpha i j} \ket{\alpha i k}_r \bra{\alpha j \ell} \otimes \ket{\overline{\alpha} i' k}_{\overline{r}} \bra{\overline{\alpha} i' \ell} : \widetilde{\mathcal{O}}_{\alpha i j} \in \mathbbm{C} \rt\},
\label{eq:GIsmart}
\ea
with $\widetilde{\mathcal{O}}_r$ in \er{eq:GIsmart} identified with $\hat{\mathcal{O}}_r \pGI$ in \er{eq:gaugedmid} under $\widetilde{\mathcal{O}}_{\alpha i j} = \hat{\mathcal{O}}_{\alpha i j}/d_\alpha$.\footnote{In a slight abuse of notation, we have referred to the matrix elements of an operator with the same symbol as the operator itself, but with irrep labels and indices such as $\a$, $i$, and $j$. We could have referred to $\hat{\cO}_{\a ij}$ as $\hat{\cO}_{r,\a ij}$ in \er{eq:gaugedmid}, but for simplicity, we will use the former.}
    \begin{proof}
        
We show this by noting $\widetilde{\mathcal{A}}_r = \hat{\mathcal{A}}_r^\Pi \pGI$ and constructing a generic operator therein.
Recall that $\{ \ket{\alpha i k}_r \}$ is a basis for $\hat{\mathcal{H}}_r$, so an operator $\hat{\mathcal{O}}_r \in \hat{\mathcal{A}}_r$ (not necessarily gauge-invariant) can be written as
\be
\hat{\mathcal{O}}_r = \sum_{\alpha \beta i j k \ell} \hat{\mathcal{O}}_{\alpha \beta i j k \ell} \ket{\alpha i k}_r \bra{\beta j \ell} \otimes \mathbbm{1}_{\overline{r}}
\ee
with some $\hat{\mathcal{O}}_{\alpha \beta i j k \ell} \in \mathbbm{C}$. Now we require $\hat{\mathcal{O}}_r \in \hat{\mathcal{A}}_r^\Pi$, so we will try to impose $\hat{\mathcal{O}}_r \pGI = \pGI \hat{\mathcal{O}}_r$. We find
\ba
\hat{\mathcal{O}}_r \pGI &= \hat{\mathcal{O}}_r \Pi_{V_r} \Pi_{V_{\overline{r}}} \Pi_{V_c} \nonumber \\
&= \( \sum_{\alpha \beta i j k \ell} \hat{\mathcal{O}}_{\alpha \beta i j k \ell} \ket{\alpha i k}_r \bra{\beta j \ell} \otimes \mathbbm{1}_{\overline{r}} \)  \Pi_{V_{\overline{r}}} \Pi_{V_c} \nonumber \\
&= \left( \sum_{\alpha \beta \gamma i j k \ell i' k'} \hat{\mathcal{O}}_{\alpha \beta i j k \ell} \ket{\alpha i k}_r \bra{\beta j \ell} \otimes \ket{\overline{\gamma} i' k'}_{\overline{r}} \bra{\overline{\gamma} i' k'}\right)  \Pi_{V_c}\nonumber \\
&= \sum_{\alpha \beta \gamma i j k \ell i' k' \ell'} \hat{\mathcal{O}}_{\alpha \beta i j k \ell} \frac{1}{d_\beta} \delta_{\beta \gamma} \delta_{\ell k'} \ket{\alpha i k}_r \bra{\beta j \ell'} \otimes \ket{\overline{\gamma} i' k'}_{\overline{r}} \bra{\overline{\gamma} i' \ell'} \nonumber \\
&= \sum_{\alpha \beta i j k \ell i' \ell'} \hat{\mathcal{O}}_{\alpha \beta i j k \ell} \frac{1}{d_\beta} \ket{\alpha i k}_r \bra{\beta j \ell'} \otimes \ket{\overline{\beta} i' \ell}_{\overline{r}} \bra{\overline{\beta} i' \ell'},
\la{eq:hOrPi}
\ea
where we have used the basis of $\hat{\mathcal{H}}_{\overline{r}}$ in going to the third line and used \er{eq:Vcaction} in going to the fourth line. We can apply the same procedure to write $\pGI \hat{\mathcal{O}}_r$ as
\ba
\pGI \hat{\mathcal{O}}_r &= \Pi_{V_c}  \sum_{\alpha \beta \gamma i j k \ell i' k'} \hat{\mathcal{O}}_{\alpha \beta i j k \ell} \ket{\alpha i k}_r \bra{\beta j \ell} \otimes \ket{\overline{\gamma} i' k'}_{\overline{r}} \bra{\overline{\gamma} i' k'}  \nonumber \\
&= \sum_{\alpha \beta \gamma i j k \ell i' k' \ell'} \hat{\mathcal{O}}_{\alpha \beta i j k \ell} \frac{1}{d_\alpha} \delta_{\alpha \gamma} \delta_{k k'} \ket{\alpha i \ell'}_r \bra{\beta j \ell} \otimes \ket{\overline{\gamma} i' \ell'}_{\overline{r}} \bra{\overline{\gamma} i' k'} \nonumber \\
&= \sum_{\alpha \beta i j k \ell i' \ell'} \hat{\mathcal{O}}_{\alpha \beta i j k \ell} \frac{1}{d_\alpha} \ket{\alpha i \ell'}_r \bra{\beta j \ell} \otimes \ket{\overline{\alpha} i' \ell'}_{\overline{r}} \bra{\overline{\alpha} i' k}.
\la{eq:PihOr}
\ea
One way to proceed is to find conditions on $\hat{\mathcal{O}}_{\alpha \beta a i j k \ell}$ such that the two expressions \er{eq:hOrPi}, \er{eq:PihOr} are equal, but doing this explicitly turns out to be slightly complicated (in cases where the multiplicities $\ol{n}_{\ol\a}$, $\ol{n}_{\ol\b}$ vanish but $n_\a$, $n_\b$ do not). Instead, we will use the equality of \er{eq:hOrPi} and \er{eq:PihOr} to directly show that $\hat{\mathcal{A}}_r^\Pi \pGI$ contains and is contained in the right-hand side of \eqref{eq:gaugedmid}, which we now define as
\be
\wtd{\cA}_r^{(3)} \eq \lt\{ \hat{\mathcal{O}}_r \pGI : \hat{\mathcal{O}}_r = \sum_{\alpha i j k} \hat{\mathcal{O}}_{\alpha i j} \ket{\alpha i k}_r \bra{\alpha j k} \otimes \mathbbm{1}_{\overline{r}},\, \hat{\mathcal{O}}_{\alpha i j} \in \mathbbm{C} \rt\}.
\la{eq:tAr3}
\ee

First, we show that $\wtd{\cA}_r^{(3)}$ defined by \er{eq:tAr3} is equal to \er{eq:GIsmart} as claimed. To see this, we simply apply \er{eq:hOrPi} to the special case of $\hat{\mathcal{O}}_r = \sum_{\alpha i j k} \hat{\mathcal{O}}_{\alpha i j} \ket{\alpha i k}_r \bra{\alpha j k} \otimes \mathbbm{1}_{\overline{r}}$ and find $\hat{\mathcal{O}}_r \Pi_{V_c}$ to be identical to $\widetilde{\mathcal{O}}_r$ in \er{eq:GIsmart} under $\widetilde{\mathcal{O}}_{\alpha i j} = \hat{\mathcal{O}}_{\alpha i j}/d_\alpha$. Moreover, applying \er{eq:PihOr} to this case yields the same operator, so we find that this special $\hat{\mathcal{O}}_r$ commutes with $\pGI$. Thus, $\wtd{\cA}_r^{(3)}$ is contained in $\hat{\mathcal{A}}_r^\Pi \pGI$.

Finally, we will show that $\hat{\mathcal{A}}_r^\Pi \pGI$ is contained in $\wtd{\cA}_r^{(3)}$. Any $\hat{\mathcal{O}}_r \pGI \in \hat{\mathcal{A}}_r^\Pi \pGI$ can be written explicitly as
\ba
\hat{\mathcal{O}}_r \pGI &= \pGI \hat{\mathcal{O}}_r \pGI = \Pi_{V_c} \sum_{\alpha \beta i j k \ell i' \ell'} \hat{\mathcal{O}}_{\alpha \beta i j k \ell} \frac{1}{d_\beta} \ket{\alpha i k}_r \bra{\beta j \ell'} \otimes \ket{\overline{\beta} i' \ell}_{\overline{r}} \bra{\overline{\beta} i' \ell'} \nonumber \\
&= \sum_{\alpha \beta i j k \ell i' k' \ell'} \hat{\mathcal{O}}_{\alpha \beta i j k \ell} \frac{1}{d_\alpha d_\beta} \delta_{\alpha \beta} \delta_{k \ell} \ket{\alpha i k'}_r \bra{\beta j \ell'} \otimes \ket{\overline{\beta} i' k'}_{\overline{r}} \bra{\overline{\beta} i' \ell'} \nonumber \\
&= \sum_{\alpha i j k \ell i' k' } \hat{\mathcal{O}}_{\alpha \alpha i j k' k'} \frac{1}{d_\alpha^2} \ket{\alpha i k}_r \bra{\alpha j \ell} \otimes \ket{\overline{\alpha} i' k}_{\overline{r}} \bra{\overline{\alpha} i' \ell},
\ea
which is identical to $\widetilde{\mathcal{O}}_r$ in \eqref{eq:GIsmart} under $\widetilde{\mathcal{O}}_{\alpha i j} = \sum_{k'} \hat{\mathcal{O}}_{\alpha \alpha i j k' k'}/d_\alpha^2$, and thus belongs to $\wtd{\cA}_r^{(3)}$.
Combining the above results, we conclude $\widetilde{\mathcal{A}}_r = \widetilde{\mathcal{A}}_r^{(3)}$. 
    \end{proof}
\end{lemma}

After all of this machinery, it is clear that one is justified in writing the algebra $\widetilde{\mathcal{A}}_r$ of gauge-invariant operators on a subregion $r$ as a restriction via $\pGI$ of the pre-gauged algebra $\cA_r$ on $r$. Crucially, however, $\widetilde{\mathcal{A}}_r$ is \textit{not} a subalgebra of $\mathcal{A}_r$, as is obvious from the nontrivial action of \eqref{eq:GIsmart} on $\hat{\mathcal{H}}_{\overline{r}}$. This is manifest from the fact that $\pGI$ is an element of $\mathcal{A}$, not of $\mathcal{A}_r$, and so the projection takes one out of the pre-gauged subregion algebra $\mathcal{A}_r$.

\subsection{The center of the algebra}
For spatial subregions the following inclusion is obvious:
\begin{equation}
    \mathcal{A}_{\overline{r}} \subseteq \left(\mathcal{A}_r\right)',
\end{equation}
as causally disconnected operators must commute. Here $\left(\mathcal{A}_r\right)'$ denotes the commutant of $\cA_r$. Haag duality is the saturation of the above bound:
\begin{equation}
    \mathcal{A}_{\overline{r}} = \left(\mathcal{A}_r\right)',
\end{equation}
that is, the commutant of the algebra of operators in a subregion is equal to the algebra of operators in the complement region.\footnote{There are counterexamples to Haag duality in quantum field theories with global or gauge symmetries; see for example \cite{Casini:2020rgj}.}

In our model, Haag duality certainly holds for the pre-gauged algebras, but does it also hold for the gauge-invariant algebras?
We will now show that it does, i.e.,
\be
\wtd{\mathcal{A}}_{\overline{r}} = \left(\wtd{\mathcal{A}}_r\right)'.
\ee

\begin{proposition}
The Hilbert space isomorphism \er{eq:hilbertISO} induces the following isomorphisms between algebras:
\be
\widetilde{\mathcal{A}}_r \cong \bigoplus_\alpha \widetilde{\mathcal{A}}_r^\alpha \otimes \mathbbm{1}^{\overline{\alpha}}_{\overline{r}}, \qqu
\widetilde{\mathcal{A}}_{\ol r} \cong \bigoplus_\alpha \mathbbm{1}_r^\alpha \otimes \widetilde{\mathcal{A}}^{\overline{\alpha}}_{\overline{r}},
\label{eq:algISO}
\ee
where $\widetilde{\mathcal{A}}_r^\alpha \eq \mathcal{B}(\widetilde{\mathcal{H}}_r^\a)$, the algebra of bounded operators on $\widetilde{\mathcal{H}}_r^\a$, and similarly we define $\widetilde{\mathcal{A}}_{\ol r}^{\ol \alpha} \eq \mathcal{B}(\widetilde{\mathcal{H}}_{\ol r}^{\ol \alpha})$. Moreover, $\mathbbm{1}_r^\alpha$, $\mathbbm{1}^{\overline{\alpha}}_{\overline{r}}$ denote the identity operators on $\widetilde{\mathcal{H}}_r^\a$, $\widetilde{\mathcal{H}}_{\ol r}^{\ol\a}$, respectively.
\begin{proof}
    Recall from \er{eq:hilbertISO} that $\widetilde{\mathcal{H}}$ is isomorphic to a direct sum of factorizing Hilbert spaces:
\be
\widetilde{\mathcal{H}} \cong \bigoplus_\alpha \widetilde{\mathcal{H}}_r^\alpha \otimes \widetilde{\mathcal{H}}_{\overline{r}}^{\overline{\alpha}} ;
\ee
where the two sides are identified under the natural isomorphism \er{eq:stateISO}, which we reproduce here:
\be
\ket{\widetilde{\psi}} = \sum_{\alpha i j k} \widetilde{\psi}_{\alpha i j} \ket{\alpha i k}_r \ket{\overline{\alpha} j k}_{\overline{r}} \qu\rightarrow\qu \sum_{\alpha i j} \sqrt{d_\alpha} \widetilde{\psi}_{\alpha i j} \ket{\alpha i}_r \ket{\overline{\alpha} j}_{\overline{r}}.
\la{eq:stateISO2}
\ee
We now apply this isomorphism to our algebra $\widetilde{\mathcal{A}}_r$. Consider a general element of $\widetilde{\mathcal{A}}_r$ defined via \eqref{eq:GIsmart}. Under \eqref{eq:stateISO2}, this element becomes 
\ba
\sum_{\alpha i i' j k \ell} \widetilde{\mathcal{O}}_{\alpha i j} \ket{\alpha i k}_r \bra{\alpha j \ell} \otimes \ket{\overline{\alpha} i' k}_{\overline{r}} \bra{\overline{\alpha} i' \ell} &\rightarrow \sum_{\alpha i i' j} d_\a \widetilde{\mathcal{O}}_{\alpha i j} \ket{\alpha i}_r \bra{\alpha j} \otimes \ket{\overline{\alpha} i' }_{\overline{r}}  \bra{\overline{\alpha} i'} \nonumber \\
&= \sum_{\alpha i j} d_\a \widetilde{\mathcal{O}}_{\alpha i j} \ket{\alpha i}_r \bra{\alpha j} \otimes \mathbbm{1}_{\overline{r}}^{\overline{\alpha}},
\ea
which is an element of $\widetilde{\mathcal{A}}_r^\alpha \otimes \mathbbm{1}^{\overline{\alpha}}_{\overline{r}}$. Thus, we have demonstrated the isomorphism for $\widetilde{\mathcal{A}}_r$ in \eqref{eq:algISO}. The isomorphism for $\widetilde{\mathcal{A}}_{\ol r}$ follows from a similar argument.
\end{proof}

\begin{corollary}
    $\widetilde{\mathcal{A}}_r$ obeys Haag duality, such that $\left(\widetilde{\mathcal{A}}_r \right)' = \widetilde{\mathcal{A}}_{\overline{r}}$, where the commutant is defined with respect to the full gauge-invariant algebra $\wtd\cA$.
    \begin{proof}
       This immediately follows from the algebra isomorphisms \er{eq:algISO} and
\be
\left( \bigoplus_\alpha \widetilde{\mathcal{A}}_r^\alpha \otimes \mathbbm{1}^{\overline{\alpha}}_{\overline{r}} \right)' = \bigoplus_\alpha \left( \widetilde{\mathcal{A}}_r^\alpha \otimes \mathbbm{1}^{\overline{\alpha}}_{\overline{r}} \right)'  =  \bigoplus_\alpha \mathbbm{1}^\alpha_r \otimes \widetilde{\mathcal{A}}_{\overline{r}}^{\overline{\alpha}}.
\ee
\end{proof}
\end{corollary}
\end{proposition}
The center of an algebra is defined to be the intersection of the algebra with its commutant. As our gauge-invariant subalgebra $\widetilde{\mathcal{A}}_r$ obeys Haag duality, the center is
\be
\widetilde{\mathcal{Z}}_r = \widetilde{\mathcal{A}}_r \cap \widetilde{\mathcal{A}}_r' = \widetilde{\mathcal{A}}_r \cap \widetilde{\mathcal{A}}_{\overline{r}}.
\ee

\begin{lemma}
The center $\widetilde{\mathcal{Z}}_r$ is
\begin{equation}
    \widetilde{\mathcal{Z}}_r = \left\{ z_\alpha \widetilde{P}^\alpha : z_\alpha \in \mathbbm{C} \right\},
\end{equation}
where $\widetilde{P}^\alpha$ are mutually orthogonal projections defined via
\begin{equation}
    \widetilde{P}^\alpha = \frac{1}{d_\alpha} \sum_{ijk \ell} \ket{\alpha i k}_r \bra{\alpha i \ell} \otimes \ket{\overline{\alpha} j k}_{\overline{r}} \bra{\overline{\alpha} j \ell}.
    \label{eq:centerprojectors}
\end{equation}
\begin{proof}
Under the algebra isomorphisms \er{eq:algISO} for $\widetilde{\mathcal{A}}_r$ and $\widetilde{\mathcal{A}}_{\overline{r}}$, we can immediately identify the center as
\be
\widetilde{\mathcal{Z}}_r \cong \bigoplus_\alpha \mathbbm{C} \left( \mathbbm{1}^\alpha_r \otimes \mathbbm{1}^{\overline{\alpha}}_{\overline{r}} \right).
\ee
That is, the center $\widetilde{\mathcal{Z}}_r$ is the direct sum of complex multiples of the identity within each superselection sector $\a$. We can write the identity in a superselection sector as
\be
\mathbbm{1}^\alpha_r \otimes \mathbbm{1}^{\overline{\alpha}}_{\overline{r}} = \sum_{i j} \ket{\alpha i}_r \bra{\alpha i} \otimes \ket{\overline{\alpha} j}_{\overline{r}} \bra{\overline{\alpha} j},
\ee
and examine the pullback of these operators under the natural isomorphism \eqref{eq:stateISO2} to find the corresponding operators in $\wtd\cA$. We obtain
\be
\mathbbm{1}^\alpha_r \otimes \mathbbm{1}^{\overline{\alpha}}_{\overline{r}} \qu\Rightarrow\qu \frac{1}{d_\alpha} \sum_{i j k \ell} \ket{\alpha i k}_r \bra{\alpha i \ell} \otimes  \ket{\overline{\alpha} j k}_{\overline{r}} \bra{\overline{\alpha} j \ell} = \widetilde{P}^{\alpha}.
\ee
We identify these operators $\widetilde{P}^{\alpha}$ as the (properly normalized) projections onto the $\alpha$ superselection sector, where we remind the reader that $\alpha$ is an irreducible representation of $G^n$. These operators can alternatively be written as
\be
\widetilde{P}^{\alpha} = \left( \hat{P}^{\alpha}_r \otimes \mathbbm{1}_{\overline{r}}\right) \pGI = \left( \mathbbm{1}_{r} \otimes \hat{P}^{\overline{\alpha}}_{\overline{r}}\right) \pGI,
\ee
where $\hat{P}^\alpha_r$ and $\hat{P}^{\overline{\alpha}}_{\overline{r}}$ are orthogonal projections in $\hat{\mathcal{H}}_r$ and $\hat{\mathcal{H}}_{\overline{r}}$, respectively:
\be
\hat{P}^\alpha_r = \sum_{i k} \ket{\alpha i k}_r \bra{\alpha i k}, \quad \hat{P}^{\overline{\alpha}}_{\overline{r}} = \sum_{ik} \ket{\overline{\alpha} i k}_{\overline{r}} \bra{\overline{\alpha} i k}.
\la{eq:centerprojectorsr}
\ee
One can show the $\widetilde{P}^\alpha$ are orthogonal and idempotent such that $\widetilde{P}^\alpha \widetilde{P}^\beta = \delta_{\alpha \beta} \widetilde{P}^\alpha$.
\end{proof}
\end{lemma}

\subsection{Traces in $\mathcal{A}_r$ and $\widetilde{\mathcal{A}}_r$}
We now define traces in our von Neumann algebras.
When an algebra is $\mathcal{B}(\mathcal{H})$ for some Hilbert space $\cH$, we can simply identify the minimal projections as projections onto a pure state in $\mathcal{H}$, and the trace is the usual trace of a square matrix. Our algebras are not always of this form; an example is $\widetilde{\mathcal{A}}_r$. Therefore, we will first identify the minimal projections, which are then used to define a normalized trace on the algebra. In particular, for $\widetilde{\mathcal{A}}_r$ our task is to find the minimal projections $\widetilde{P}_r$ in $\widetilde{\mathcal{A}}_r$ and use them to define a ``rescaled'' trace $\widetilde{\Tr}_r$ which satisfies
\be
\widetilde{\Tr}_r \widetilde{P}_r = 1.
\ee

Let us first discuss the case that we understand well: that of $\mathcal{A}_r$, and by extension $\hat{\mathcal{A}}_r$. As $\mathcal{A}_r = \mathcal{B}(\mathcal{H}_r)$, the minimal projections are projections onto a pure state in $\mathcal{H}_r$, and we define the trace $\Tr_r$ in $\mathcal{A}_r$ such that the minimal projections have trace $1$.
As $\hat{\mathcal{A}}_r = \mathcal{B}(\hat{\mathcal{H}}_r)$, we proceed similarly.
Recall that the basis states of $\hat{\mathcal{H}}_r$ are $\{ \ket{\alpha i k}_r\}$, and so we define the trace $\hat{\Tr}_r$ in $\hat{\mathcal{A}}_r$ via
\be
\hat{\Tr}_r \ket{\alpha i k}_r \bra{\alpha i k} = 1.
\ee
As the minimal projections in $\hat{\mathcal{A}}_r$ are also minimal projections in $\mathcal{A}_r$, the two traces agree (on $\hat{\mathcal{A}}_r$):
\be
\Tr_r = \hat{\Tr}_r,
\ee
so we will use only $\Tr_r$ (not $\hat{\Tr}_r$) moving forward.

Now consider $\widetilde{\mathcal{A}}_r$. Although $\widetilde{\mathcal{A}}_r$ is not the algebra of all bounded operators on a Hilbert space, the algebra isomorphism \eqref{eq:algISO} shows that we can write it as a direct sum of algebras for which we can easily identify minimal projections. In particular, the pullback of minimal projections onto pure states $\ket{\alpha i}_r \in \widetilde{\mathcal{H}}_r^\alpha$ under the natural isomorphism \er{eq:stateISO2} gives minimal projections in $\widetilde{\mathcal{A}}_r$.
Thus, we write these minimal projections $\widetilde{P}^{\alpha i}_r \in \widetilde{\mathcal{A}}_r$ as 
\be
\widetilde{P}^{\alpha i}_r = \frac{1}{d_\alpha} \sum_{j k \ell} \ket{\alpha i k}_r \bra{\alpha i \ell} \otimes \ket{\overline{\alpha} j k}_{\overline{r}} \bra{\overline{\alpha} j \ell}
\la{eq:tPairdef}
\ee
for all non-empty sectors $\a$, defined as those with nonzero $n_\a$, $\ol{n}_{\ol\a}$. If $n_\a$ vanishes, the index $i$ above has an empty range, and if $\ol{n}_{\ol\a}$ vanishes, $\widetilde{P}^{\alpha i}_r$ vanishes due to the empty sum over $j$ in \er{eq:tPairdef}.

We can alternatively write $\widetilde{P}^{\alpha i}_r$ as
\be
\widetilde{P}^{\alpha i}_r = \hat{P}^{\alpha i}_r \pGI,
\ee
where the projections $\hat{P}^{\alpha i}_r$ are defined similarly to \eqref{eq:centerprojectorsr}:
\be
\hat{P}^{\alpha i}_r \equiv \sum_k \ket{\alpha i k} \bra{\alpha i k}_r \otimes \mathbbm{1}_{\overline{r}}.
\ee

Although we already argued that $\widetilde{P}^{\alpha i}_r$ are minimal projections using the natural isomorphism, we now show it more directly.

\begin{lemma}
    The projections $\widetilde{P}_r^{\alpha i}$ (for non-empty sectors $\a$) are minimal projections in $\widetilde{\mathcal{A}}_r$.
    \begin{proof}
    We recall that minimal projections are nonzero and have the property that any subprojection $\widetilde{Q}_r$ of $\widetilde{P}^{\alpha i}_r$ is either zero or $\widetilde{P}^{\alpha i}_r$. As an element of $\widetilde{\mathcal{A}}_r$, $\widetilde{Q}_r$ must be of the form
\be
\widetilde{Q}_r = \sum_{\beta j j' k} \hat{Q}_{\beta j j'} \left( \ket{\beta j k}_r \bra{\beta j' k} \otimes \mathbbm{1}_{\overline{r}} \right) \pGI
\ee
with complex coefficients $\hat{Q}_{\beta j j'}$.
The subprojection $\widetilde{Q}_r$ is left fixed under conjugation via $\widetilde{P}_r^{\alpha i}$, so we have
\be
\widetilde{Q}_r = \widetilde{P}_r^{\alpha i} \widetilde{Q}_r \widetilde{P}_r^{\alpha i} = \sum_k \hat{Q}_{\alpha ii} \left( \ket{\alpha i k}_r \bra{\alpha i k} \otimes \mathbbm{1}_{\overline{r}} \right) \pGI = \hat{Q}_{\alpha ii} \widetilde{P}^{\alpha i}_r.
\ee
Additionally imposing $\widetilde{Q}_r^2 = \widetilde{Q}_r$, we find
\be
\widetilde{Q}_r^2 = \hat{Q}_{\alpha ii}^2 \widetilde{P}^{\alpha i}_r = \hat{Q}_{\alpha ii} \widetilde{P}^{\alpha i}_r.
\ee
Unless $\widetilde{Q}_r$ is zero, we obtain $\hat{Q}_{\alpha ii} = 1$ and thus $\widetilde{Q}_r = \widetilde{P}_r^{\alpha i}$. So $\widetilde{P}_r^{\alpha i}$ (for a non-empty sector $\a$) is indeed a minimal projection. 
    \end{proof}
\end{lemma}

Therefore, we define the trace $\widetilde{\Tr}_r$ in $\widetilde{\mathcal{A}}_r$ by imposing
\be
\widetilde{\Tr}_r \widetilde{P}_r^{\alpha i} = 1
\ee
for every non-empty sector $\a$ and every $i=1,\cd,n_\a$.

How do we understand this trace acting on a general operator in $\widetilde{\mathcal{A}}_r$? Such an operator can always be written in the form \eqref{eq:gaugedmid}:
\be
\widetilde{\mathcal{O}}_r = \hat{\mathcal{O}}_r \pGI, \quad \hat{\mathcal{O}}_r = \sum_{\alpha i j k} \hat{\mathcal{O}}_{\alpha i j} \ket{\alpha i k}_r \bra{\alpha j k} \otimes \mathbbm{1}_{\overline{r}}.
\ee
Taking the trace using $\widetilde{\Tr}_r$, we find
\be
\widetilde{\Tr}_r \widetilde{\mathcal{O}}_r = \widetilde{\Tr}_r \sum_{\alpha i} \hat{\mathcal{O}}_{\alpha ii} \widetilde{P}_r^{\alpha i} = \sum_{\alpha i} \hat{\mathcal{O}}_{\alpha ii}.
\label{eq:tr1}
\ee
If we were to take the trace $\Tr_r$ of the corresponding $\hat{\mathcal{O}}_r$, we would instead find
\be
\Tr_r \hat{\mathcal{O}}_r = \Tr \sum_{\alpha i} \hat{\mathcal{O}}_{\alpha ii} \hat{P}_r^{\alpha i} = \sum_{\alpha i} d_\alpha \hat{\mathcal{O}}_{\alpha ii}.
\label{eq:tr2}
\ee
Thus, it is tempting to relate the trace $\widetilde{\Tr}_r$ to $\Tr_r$ using an appropriate rescaling by $1/d_\a$ in each sector.
A more precise version of this statement is the following: for any operator $\widetilde{\mathcal{O}}_r^\alpha \in \widetilde{\mathcal{A}}_r$ that acts only in the $\alpha$ sector such that it can be written as
\be
\widetilde{\mathcal{O}}_r^\alpha = \hat{\mathcal{O}}_r^\alpha
\pGI, \quad \hat{\mathcal{O}}_r^\alpha = \sum_{ijk} \hat{\mathcal{O}}_{\alpha i j} \ket{\alpha i k}_r \bra{\alpha j k} \otimes \mathbbm{1}_{\overline{r}},
\ee
i.e., with no sum over $\alpha$, the two traces are related by
\be
\widetilde{\Tr}_r \widetilde{\mathcal{O}}_r^\alpha  = \frac{1}{d_\alpha}\Tr_r \hat{\mathcal{O}}_r^\alpha.
\label{eq:tracerelation}
\ee
Summing both sides over $\a$ recovers \eqref{eq:tr1} and \eqref{eq:tr2}.

\subsection{Reduced states}

Our ultimate goal is to relate the von Neumann entropies for the same gauge-invariant state $\widetilde{\rho}$ in $\widetilde{\mathcal{H}}$ on two different subalgebras: $\mathcal{A}_r$ and $\widetilde{\mathcal{A}}_r$.

The first thing to note is that, when we consider the full graph (instead of restricting to a subregion $r$), we have $\cA=\cB(\cH)$ and $\wtd\cA=\cB(\wtd\cH)$ where $\wtd\cH$ is a subspace of $\cH$, so minimal projections in $\wtd\cA$ are also minimal projections in $\cA$, and the trace $\widetilde{\Tr}$ in $\wtd\cA$ therefore agrees with the trace $\Tr$ in $\cA$ when acting on gauge-invariant states. Hence, a gauge-invariant state $\wtd\r$ on the full graph that is properly normalized under the $\wtd\Tr$ trace is also properly normalized under the $\Tr$ trace, and can therefore be viewed as a properly normalized state $\r = \wtd\r$ in $\cH$ (albeit a special one). Thus, we will use only $\r$ (not $\wtd\r$) for notational simplicity in the following discussions. We should still remember that $\r$ is a special state that belongs to $\wtd\cA$.

The above statements do not hold for reduced states on subregions. In particular, we need to distinguish a properly normalized state $\r_r$ in $\cA_r$ from a properly normalized $\wtd\r_r$ in $\wtd\cA_r$. Now we derive the relation between these two states.

Recall that to find $S(\rho, \mathcal{A}_r)$ for a general subalgebra $\mathcal{A}_r \subset \mathcal{A}$, we need to find a reduced state $\rho_r \in \mathcal{A}_r$ satisfying
\be
\Tr_r (\rho_r \mathcal{O}_r) = \Tr (\rho \mathcal{O}_r) 
\ee
for all $\mathcal{O}_r \in \mathcal{A}_r$. For our particular $\mathcal{A}_r$ (the pre-gauged algebra on $r$), the answer is, of course, $\rho_r = \Tr_{\overline{r}} \rho$.

Now we work out the reduced state in the subalgebra $\wtd\cA_r$.

\begin{lemma}
The reduced state $\widetilde{\rho}_r \in \widetilde{\mathcal{A}}_r$ satisfying
\be
\widetilde{\Tr}_r ( \widetilde{\rho}_r \widetilde{\mathcal{O}}_r ) = 
\Tr (\rho \widetilde{\mathcal{O}}_r )
\label{eq:rhocond}
\ee
for all $\widetilde{\mathcal{O}}_r \in \widetilde{\mathcal{A}}_r$ is of the form 
\begin{equation}
    \widetilde{\rho}_r = \hat{\rho}_r \pGI, \quad \hat{\rho}_r = \sum_{\alpha i j k} \hat{\rho}_{\alpha i j} \ket{\alpha i k}_r \bra{\alpha j k} \otimes \mathbbm{1}_{\overline{r}},
    \la{eq:rhoGI}
\end{equation}
with $\hat{\rho}_{\alpha i j} = d_\alpha \rho_{\alpha i j}$, where $\rho_{\alpha i j}$ is defined by
\be
\rho_r = \sum_{\alpha i j k} \rho_{\alpha i j } \ket{\alpha i k}_r \bra{\alpha j k}.
\la{eq:rhorexpand}
\ee
\begin{proof}
    
A general gauge-invariant state $\rho \in \widetilde{\mathcal{A}}$ can be written as
\be
\rho = \sum_{\alpha \beta ijk  i' j' k' } \rho_{\alpha \beta i i' j j'}  \ket{\alpha  i k}_r \ket{\overline{\alpha} j k}_{\overline{r}} \bra{\beta i' k'}_r \bra{\overline{\beta} j' k'}_{\overline{r}} 
\ee
using the basis states for $\widetilde{\mathcal{H}}$. Tracing over $\overline{r}$, we find
\ba
\rho_r = \Tr_{\overline{r}} \rho
&= \sum_{\alpha \beta i j k i' j' k'} \rho_{\alpha \beta i i' j j' } \expval{\overline{\beta} j' k' | \overline{\alpha} j k}_{\overline{r}}  \ket{\alpha i' k'}_r \bra{\beta i k} \nonumber \\
&= \sum_{\alpha i i' j k} \rho_{\alpha \alpha i i' j j } \ket{\alpha i k}_r \bra{\alpha i' k}.
\ea
This verifies \er{eq:rhorexpand} and determines $\r_{a ij}$.

Now recall that as an element of $\widetilde{\mathcal{A}}_r$, $\wtd\r_r$ must be of the form \er{eq:rhoGI} with some complex coefficients $\hat{\r}_{\a ij}$. It remains to determine what they are from \eqref{eq:rhocond}.
In order to impose it, we define the following basis for $\widetilde{\mathcal{A}}_r$:
\be
\widetilde{\mathcal{O}}^{\alpha i j}_r = \hat{\mathcal{O}}^{\alpha i j}_r \pGI, \quad \hat{\mathcal{O}}^{\alpha i j}_r = \sum_{k} \ket{\alpha i k}_r \bra{\alpha j k} \otimes \mathbbm{1}_{\overline{r}},
\ee
such that we can rewrite the reduced gauge-invariant density matrix as
\be
\widetilde{\rho}_r = \sum_{\alpha i j} \hat{\rho}_{\alpha i j} \widetilde{\mathcal{O}}_{\alpha i j}.
\ee
Note that the basis elements $\widetilde{\mathcal{O}}^{\alpha i j}_r$ and their corresponding basis elements $\hat{\mathcal{O}}^{\alpha i j}_r \in \hat{\mathcal{A}}_r$ obey the following relations:
\be
\widetilde{\mathcal{O}}^{\alpha i j}_r \widetilde{\mathcal{O}}^{\beta i' j'}_r = \delta_{\alpha \beta} \delta_{i' j} \widetilde{\mathcal{O}}^{\alpha i j'}_r, \quad \widetilde{\Tr}_r \widetilde{\mathcal{O}}^{\alpha i j}_r = \delta_{ij}
\ee
\be
\hat{\mathcal{O}}^{\alpha i j}_r \hat{\mathcal{O}}^{\beta i' j'}_r = \delta_{\alpha \beta} \delta_{i' j} \hat{\mathcal{O}}^{\alpha i j'}_r, \quad \Tr_r \hat{\mathcal{O}}^{\alpha i j}_r = d_\alpha \delta_{ij}.
\ee
From these relations we can check both sides of \eqref{eq:rhocond} for $\widetilde{\mathcal{O}}_r$ set to one of the basis elements $\widetilde{\mathcal{O}}^{\alpha i j}_r$. The trace in the gauge-invariant algebra becomes
\be
\widetilde{\Tr}_r \left( \widetilde{\rho}_r \widetilde{\mathcal{O}}^{\alpha i j}_r \right) = \widetilde{\Tr}_r \left( \sum_{\beta i' j'} \hat{\rho}_{\beta i' j'} \widetilde{\mathcal{O}}^{\beta i' j'}_r \widetilde{\mathcal{O}}^{\alpha i j} \right) = \sum_{\beta i' j'} \hat{\rho}_{\beta i' j'} \delta_{\alpha \beta} \delta_{i' j} \delta_{i j'} = \hat{\rho}_{\alpha j i}.
\ee
We need to equate this with the trace in the pre-gauged algebra, which we begin to evaluate by simplifying to the trace in $\mathcal{A}_r$. We have
\be
\Tr (\rho \widetilde{\mathcal{O}}^{\alpha i j}_r ) = \Tr (\rho \hat{\mathcal{O}}^{\alpha i j}_r \pGI ) = \Tr (\pGI \rho \hat{\mathcal{O}}^{\alpha i j}_r ) = \Tr ( \rho \hat{\mathcal{O}}^{\alpha i j}_r ) = \Tr_r (\rho_r \hat{\mathcal{O}}^{\alpha i j}_r ),
\label{eq:ohatrelations}
\ee
where we have used the cyclicity of the trace, the gauge invariance of $\rho$, and the fact that $\hat{\mathcal{O}}^{\alpha i j}_r \in \mathcal{A}_r$. We further simplify this and obtain
\be
\Tr_r (\rho_r \hat{\mathcal{O}}^{\alpha i j}_r ) = \Tr_r \left( \sum_{\beta i' j'} \rho_{\beta i' j'} \hat{\mathcal{O}}_r^{\beta i' j'} \hat{\mathcal{O}}_r^{\alpha i j} \right) = \sum_{\beta i' j'} d_\alpha \rho_{\beta i' j'} \delta_{\alpha \beta} \delta_{i' j} \delta_{i j'} = d_\alpha \rho_{\alpha j i}.
\ee
Thus we identify the reduced density matrix $\widetilde{\rho}_r \in \widetilde{\mathcal{A}}_r$ as a density matrix of the form \eqref{eq:rhoGI} with
\be
\hat{\rho}_{\alpha i j} = d_\alpha \rho_{\alpha i j}.
\ee
\end{proof}
\end{lemma}

\section{Entropies in the Gauged Random Tensor Network}
\label{sec:entropy}

Having written down the reduced states in $\mathcal{A}_r$ and $\widetilde{\mathcal{A}}_r$, we are now ready to compute the von Neumann entropies with respect to the two algebras. As we will see, the difference between the two entropies in the gauged random tensor network is precisely accounted for by an additional contribution to the area operator in the non-trivial center $\widetilde{\mathcal{Z}}_r$.

\subsection{Entanglement entropy}
\label{subsec:EE}
From \er{eq:rhoGI} and \er{eq:rhorexpand}, we proceed by defining the reduced states projected onto a superselection sector $\alpha$:
\be
\rho^\alpha_r = \sum_{ijk} \rho_{\alpha ij} |\alpha ik\>_r\< \alpha jk|, \qqu \hat{\rho}^\alpha_r = \sum_{ijk} \hat{\rho}_{\alpha ij} |\alpha ik\>_r\<\alpha jk| = d_\alpha \rho_r^\alpha.
\ee
Note that these density matrices are not properly normalized with respect to their appropriate traces. The reduced states \er{eq:rhoGI} and \er{eq:rhorexpand} can be written as a direct sum over representations:
\be
\rho_r = \bigoplus_\alpha \rho^\alpha_r, \qqu \widetilde{\rho}_r = \bigoplus_\alpha \hat{\rho}^\alpha_r \pGI.
\ee
Furthermore, functions of the reduced states are superselected in the same way. In particular,
\be
\rho_r \log \rho_r = \bigoplus_\alpha \rho^\alpha_r \log \rho^\alpha_r, \qqu \widetilde{\rho}_r \log \widetilde{\rho}_r = \bigoplus_\alpha (\hat{\rho}^\alpha_r \log \hat{\rho}^\alpha_r) \pGI,
\ee
where we used the fact that $[\hat{\rho}^\alpha_r, \pGI] = 0$.

We are now ready to compute the subregion entropies (in the bulk). The von Neumann entropy of $\r$ with respect to $\mathcal{A}_r$ is simply given by
\be
S(\rho, \mathcal{A}_r) = -\Tr_r \rho_r \log \rho_r = - \sum_\alpha \Tr_r \rho_r^\alpha \log \rho_r^\alpha.
\ee
On the other hand, using the relation between the traces \eqref{eq:tracerelation}, we can write the von Neumann entropy with respect to $\widetilde{\mathcal{A}}_r$ as
\be
S(\rho, \widetilde{\mathcal{A}}_r) = -\widetilde{\Tr}_r \widetilde{\rho}_r \log \widetilde{\rho}_r = - \sum_\alpha d_\alpha^{-1} \Tr_r \hat{\rho}_r^\alpha \log \hat{\rho}_r^\alpha.
\ee
Using $\hat{\rho}^\alpha_r = d_\alpha \rho^\alpha_r$ and $\Tr_r \rho^\alpha_r = \widetilde{\Tr}_r \widetilde{\rho}^\alpha_r$, we can rewrite each term in the sum as
\ba
d_\alpha^{-1} \Tr_r \hat{\rho}^\alpha_r \log \hat{\rho}^\alpha_r &= \Tr_r \rho^\alpha_r \log \rho^\alpha_r + \Tr_r \rho^\alpha_r \log d_\alpha \nonumber \\
&= \Tr_r \rho^\alpha_r \log \rho^\alpha_r + \widetilde{\Tr}_r \widetilde{\rho}^\alpha_r \log d_\alpha.
\ea
The von Neumann entropy with respect to $\wtd\cA_r$ can thus be written as
\ba
S(\rho,\widetilde{\mathcal{A}}_r) &= - \sum_\alpha \left( \Tr_r \rho_r^\alpha \log \rho_r^\alpha + \widetilde{\Tr}_r \widetilde{\rho}_r^\alpha \log d_\alpha \right) \nonumber \\
&= S(\rho,\mathcal{A}_r) - \widetilde{\Tr}_r \left( \widetilde{\rho}_r \Delta \widetilde{A} \right) ,
\ea
where we have defined a new ``extra area operator'' via
\be
\Delta \widetilde{A} \equiv \bigoplus_\alpha \widetilde{P}^\alpha \log d_\alpha .
\ee
The projections $\widetilde{P}^\alpha$ are precisely the projections \eqref{eq:centerprojectors} which generate the center $\widetilde{\mathcal{Z}}_r$, so $\Delta \widetilde{A}$ is manifestly an operator in the center. 

We have now arrived at our final relation between the entropies with respect to $\mathcal{A}_r$ and $\widetilde{\mathcal{A}}_r$,
\be
S(\rho,\mathcal{A}_r) = S(\rho,\widetilde{\mathcal{A}}_r) + \widetilde{\Tr}_r \left( \widetilde{\rho}_r \Delta \widetilde{A} \right) ,
\label{eq:entropyrelation}
\ee
which we now use in our two-layer gauged RTN defined in Section \ref{sec:RTN}. In particular, we would like to derive an FLM formula relating the boundary entropy with the gauged bulk entropy $S(\rho,\widetilde{\mathcal{A}}_r)$.
Recall that when we feed any bulk state $\rho$ in the pre-gauged algebra $\mathcal{A}$ into the RTN, the entropy $S(R)$ of the resulting boundary state on a boundary subregion $R$ satisfies an FLM formula:
\be
S(R) = \abs{\gamma_R} \log D + S(\rho,\mathcal{A}_r),
\ee
where the bulk subregion $r$ is chosen to be the entanglement wedge between $R$ and its minimal surface $\g_R$.
Now specializing to a gauge-invariant bulk state $\r \in \wtd{\cA}$ and using \eqref{eq:entropyrelation}, we find that the boundary entropy can now be written as a new FLM formula:
\be
S(R) = \widetilde{\Tr}_r \Big( \widetilde{\rho}_r \widetilde{A} \Big) + S\(\rho, \widetilde{\mathcal{A}}_r \),
\ee
where the full area operator $\widetilde{A}$ is
\be
\widetilde{A} \, = \, \abs{\gamma_R} \log D \, + \, \bigoplus_\alpha \widetilde{P}^\alpha \log d_{\alpha} \, = \, \abs{\gamma_R} \log D \,\,\, + \!\! \bigoplus_{\alpha_1, \cdots, \alpha_n} \!\!\! \widetilde{P}^{(\alpha_1, \cdots, \alpha_n)} \sum_{i = 1}^n \log d_{\alpha_i}.
\ee
Again, we sum over all irreps $\a=(\a_1,\cd,\a_n)$ of $G^n$ acting on the cut, although some $\a$ sectors may be emtpy (i.e., $n_\a$ or ${\ol n}_{\ol \a}$ is zero) in which case $\widetilde{P}^\alpha$ vanishes.

This is our main result. We note that this area operator looks like what arises in a superposition of a stack of standard RTNs with probabilities determined by the projections $\widetilde{P}^\alpha$ and with bond dimensions augmented by $d_{\alpha_i}$.

\subsection{R\'enyi entropy and R\'enyi mutual information}

As discussed in Section \ref{sec:RTN}, one can modify the entanglement structure of the links in the standard RTN to obtain a non-flat R\'enyi spectrum for boundary states. However, this is not enough to reproduce the properties of holographic R\'enyi entropies on general boundary subregions. In particular, it fails to account for the lack of backreaction, displayed in the tensor network as a lack of (R\'enyi) correlation between disconnected boundary subregions when the RT surface is in a disconnected phase. This problem becomes clear when one calculates the R\'enyi mutual information between two such boundary subregions $R_1$ and $R_2$, defined as\footnote{The R\'enyi index $n$ should not be confused with the number of vertices on the cut $n = |V_c|$.}
\begin{equation}
    I_n(R_1 : R_2) \equiv S_n(R_1) + S_n(R_2) - S_n(R_1 \cup R_2).
\end{equation}
As the area operator in the original RTN is a $c$-number, using \er{eq:RTNentropy} we find that the area operator contribution cancels out in $I_n(R_1 : R_2)$ for all $n$ (as long as the minimal surface $\g_R$ is in a disconnected phase), leaving the boundary mutual information equal to the bulk mutual information:
\begin{equation}
I_n(R_1: R_2) = I_n(r_1: r_2, \mathcal{A}_{r_1 r_2}).
\la{eq:bdyRenMI}
\end{equation}
This implies that, if one wants a contribution to the R\'enyi mutual information of the same order as the area, that is $\mathcal{O}(\log D)$, one must input by hand a highly entangled bulk state. Doing this is unsatisfying and quite arbitrary.

\begin{figure}
    \centering
    \includegraphics[width=0.8\textwidth]{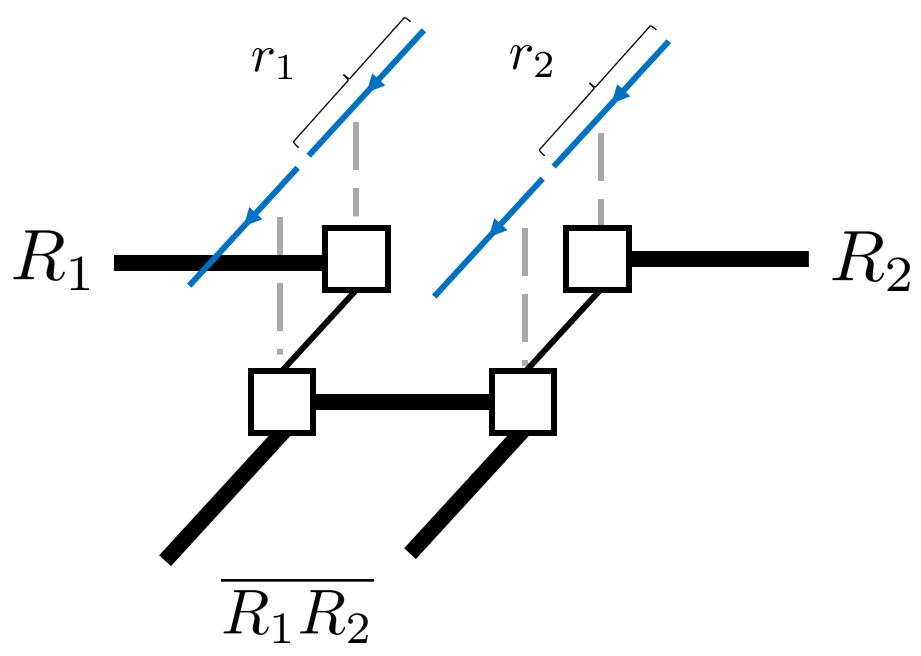}
    \caption{A simple gauged RTN in which we compute the R\'enyi mutual information between $R_1$ and $R_2$. The input from the top layer lives on four edges of a disconnected graph $G$, as we choose to have no matter on any of the vertices. In the bottom layer, the thick legs have a bond dimension much larger than that of the thin legs, such that the minimal surfaces for the three boundary regions $R_1$, $R_2$, and $R_1 \cup R_2$ only involve the light internal legs. Consequently, the associated bulk regions will be $r_1$, $r_2$, and $r_1 \cup r_2$.} 
    \label{fig:mutualnetwork}
\end{figure}

We will now see that our gauged RTN solves this problem in a natural way, due to our nontrivial area operator.
In general, the presence of a nontrivial area operator will lead to a nontrivial, $n$-dependent boundary R\'enyi mutual information, even for states with vanishing bulk R\'enyi mutual information.

To see how this is realized in the gauged RTN, we will study a simple example shown in Figure \ref{fig:mutualnetwork}, where the top layer is disconnected but the bottom layer is connected.\footnote{This connection is unnecessary to prove our point, as the internal leg connecting $r_1$ and $r_2$ never contributes to the area term, but it is more intuitively satisfying to discuss a connected spatial slice for the purposes of demonstrating backreaction.} We allow the bond dimensions in the bottom layer to be different for different links, and in fact design them so that the minimal surfaces associated with $R_1$, $R_2$, and their union $R_1 \cup R_2$ are fixed as we vary the R\'enyi index $n$ at $\mathcal{O}(1)$ values. We will feed in a gauge-invariant bulk state $\r$ with the following reduced state on $r_1 \cup r_2$:
\begin{equation}
    \rho_{r_1 r_2} = \sum_{\alpha \beta} (d_\alpha d_\beta)^{-1} P(\alpha, \beta) \sum_{k \ell} \ket{\alpha i k}_{r_1} \ket{\beta j \ell}_{r_2} \bra{\alpha i k}_{r_1} \bra{\beta j \ell}_{r_2},
\end{equation}
for some particular choice of $i$, $j$.
This state has classical correlations between $r_1$ and $r_2$ as described by a probability distribution $P(\alpha, \beta)$, but has no quantum correlations.
For simplicity, we consider the following distribution $P(\alpha, \beta)$ that has support on only two superselection sectors $\a_1$, $\a_2$ on $r_1$ and only two sectors $\b_1$, $\b_2$ on $r_2$:
\begin{equation}
    P(\alpha_1, \beta_1) = p, \quad P(\alpha_2, \beta_1) = P(\alpha_1, \beta_2) = p', \quad P(\alpha_2, \beta_2) = p'',
\end{equation}
subject to the constraint $p+2p'+p''=1$.

The R\'enyi entropy of $\rho$ in the pre-gauged algebra $\mathcal{A}_{r_1 r_2}$ is defined as
\be
    S_n(\rho, \mathcal{A}_{r_1 r_2}) \eq \frac{1}{1-n} \log \(\Tr_{r_1 r_2} \r_{r_1 r_2}^n\).
\ee
Using our $\rho_{r_1 r_2}$, we find
\begin{align}
    S_n(\rho, \mathcal{A}_{r_1 r_2}) = \frac{1}{1-n} \log &\bigg( d_{\alpha_1} d_{\beta_1} \left( \frac{p}{d_{\alpha_1} d_{\beta_1}}\right)^n + d_{\alpha_2} d_{\beta_1} \left( \frac{p'}{d_{\alpha_2} d_{\beta_1}}\right)^n \nonumber  \\
    &+ d_{\alpha_1} d_{\beta_2} \left( \frac{p'}{d_{\alpha_1} d_{\beta_2}}\right)^n + d_{\alpha_2} d_{\beta_2} \left( \frac{p''}{d_{\alpha_2} d_{\beta_2}}\right)^n \bigg).
\end{align}
We can also compute the reduced density matrices on $r_1$ and $r_2$, as well as their corresponding R\'enyi entropies in the pre-gauged algebra. We find the reduced density matrices to be
\begin{align}
    \rho_{r_1} &= \sum_{k = 1}^{d_{\alpha_1}} d_{\alpha_1}^{-1} (p + p') \ket{\alpha_1 i k}_{r_1} \bra{\alpha_1 i k} + \sum_{k' = 1}^{d_{\alpha_2}} d_{\alpha_2}^{-1} (p' + p'') \ket{\alpha_2 i k'}_{r_1} \bra{\alpha_2 i k'}, \nonumber \\
    \rho_{r_2} &= \sum_{k = 1}^{d_{\beta_1}} d_{\beta_1}^{-1} (p + p') \ket{\beta_1 j k}_{r_2} \bra{\beta_1 j k} + \sum_{k' = 1}^{d_{\beta_2}} d_{\beta_2}^{-1} (p' + p'') \ket{\beta_2 j k'}_{r_2} \bra{\beta_2 j k'},
\end{align}
and the bulk R\'enyi entropies are
\begin{align}
    S_n(\rho, \mathcal{A}_{r_1}) &= \frac{1}{1-n} \log \left( d_{\alpha_1} \left( \frac{p + p'}{d_{\alpha_1}}\right)^n + d_{\alpha_2} \left( \frac{p' + p''}{d_{\alpha_2}}\right)^n \right) \nonumber \\
    S_n(\rho, \mathcal{A}_{r_2}) &= \frac{1}{1-n} \log \left( d_{\beta_1} \left( \frac{p + p'}{d_{\beta_1}}\right)^n + d_{\beta_2} \left( \frac{p' + p''}{d_{\beta_2}}\right)^n \right).
\end{align}
In the gauge-invariant algebra, the dependence on irrep dimensions drops out and the R\'enyi entropies become purely Shannon terms:
\begin{align}
    S_n(\rho, \widetilde{\mathcal{A}}_{r_1}) = S_n(\rho, \widetilde{\mathcal{A}}_{r_2}) &= \frac{1}{1-n} \log \left( \left( p + p'\right)^n +  \left( p' + p'' \right)^n \right) \nonumber \\
    S_n(\rho, \widetilde{\mathcal{A}}_{r_1 r_2}) &= \frac{1}{1-n} \log \left( p^n + 2(p')^n + (p'')^n\right),
\end{align}
which we choose to be parametrically suppressed relative to the R\'enyi entropies in the pre-gauged algebra.

When the sum inside the logarithm is dominated by one term, we can approximate it using
\begin{equation}\label{eq:logmin}
    \log \Big( {\textstyle \sum_i} x_i \Big) \approx \log \Big( \max_i \left\{ x_i \right\} \Big).
\end{equation}
To simplify our calculation, we will enter a parameter regime where all three (pre-gauged) R\'enyi entropies satisfy the approximation above and have phase transitions.

First consider $S_n(\rho, \mathcal{A}_{r_1})$. We take $d_{\alpha_1} > d_{\alpha_2}$. The two terms in the sum are equal at some critical $n_*$, given by 
\begin{equation}
    \left( \frac{p + p'}{p' + p''} \right)^{n_*} = \left( \frac{d_{\alpha_1}}{d_{\alpha_2}}\right)^{n_* - 1} \qu\Rightarrow\qu \frac{n_*}{n_* - 1} = \frac{\log \left( \frac{d_{\alpha_1}}{d_{\alpha_2}} \right)}{\log \left( \frac{p + p'}{p' + p''}\right)}.
\end{equation}
Thus, in order to have a phase transition at $n_* > 1$ we require
\begin{equation}
\log \left( \frac{d_{\alpha_1}}{d_{\alpha_2}} \right) > \log \left( \frac{p + p'}{p' + p''}\right).
\end{equation}
The width of this transition is controlled by the corrections to \eqref{eq:logmin}. This depends on the curvature of $S_n(\rho, \mathcal{A}_{r_1})$ at $n_*$; explicitly we can diagnose this with the following quantity:
\begin{equation}
    \left. \frac{d^2}{dn^2} (1-n) S_n(\rho, \mathcal{A}_{r_1}) \right \vert_{n = n_*} = \frac{1}{4} \left(\log \frac{d_{\alpha_1} (p'+p'')}{d_{\alpha_2} (p+p')} \right)^2.
\end{equation}
For fixed $n_*$, this quantity increases with increasing $d_{\alpha_1}/d_{\alpha_2}$, so we should make this ratio large for a sharp transition. A simple way to ensure the previous conditions is the following:
\begin{equation}
    \frac{d_{\alpha_1}}{d_{\alpha_2}} \equiv q \gg 1,\qu p \gg p',\qu p' \gg p''.
\end{equation}
Furthermore, we impose
\begin{equation}
    \frac{d_{\alpha_1}}{d_{\alpha_2}} = \frac{d_{\beta_1}}{d_{\beta_2}} = q,
\end{equation}
which forces the phase transitions in $S_n(\rho, \mathcal{A}_{r_1})$ and $S_n(\rho, \mathcal{A}_{r_2})$ to occur at the same critical $n_*$.

Now let us examine the phase transition in $S_n(\rho, \mathcal{A}_{r_1 r_2})$. In the limit of sharp transitions we have
\begin{equation}
    S_n(\rho, \mathcal{A}_{r_1 r_2}) \approx \frac{1}{1-n} \log \left( \max \left\{ \frac{p^n}{(d_{\alpha_1} d_{\beta_1})^{n-1}}, \frac{(p')^n}{(d_{\alpha_2} d_{\beta_1})^{n-1}}, \frac{(p')^n}{(d_{\alpha_1} d_{\beta_2})^{n-1}},\frac{(p'')^n}{(d_{\alpha_2} d_{\beta_2})^{n-1}} \right\} \right).
\end{equation}
For simplicity, we will choose
\begin{equation}
    \frac{p}{p'} > \frac{p'}{p''} \gg 1.
\end{equation}
In this case, we find that $S_n(\rho, \mathcal{A}_{r_1 r_2})$ has a phase transition occurring at a critical $n_c$ determined by
\begin{equation}
    \frac{n_c}{n_c - 1} = \frac{\log (q^2)}{\log \left( \frac{p}{p''} \right)} = \frac{\log (q^2)}{\log \left( \frac{p}{p'} \frac{p'}{p''} \right)}
\end{equation}
which satisfies $1<n_c<n_*$.

\begin{figure}
    \centering
    \includegraphics[width=\textwidth]{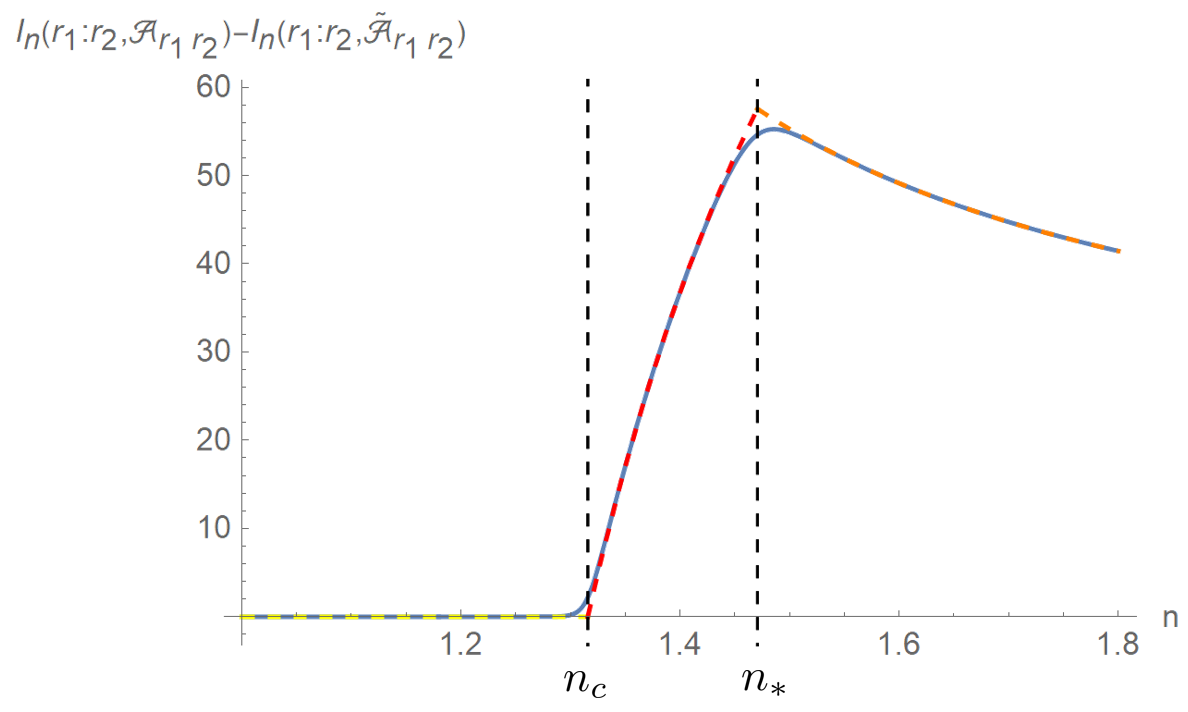}
    \caption{Phase transitions in the R\'enyi mutual information. Here we set $q = 10^{50}$, $p' = 10^{-16}$, and $p''=10^{-24}$. We plot the dominant contribution to the R\'enyi mutual information in the three phases (dashed) as well as the fully analytic interpolating function (solid).}
    \label{fig:mutual}
\end{figure}

We now combine the above results to find the (pre-gauged) R\'enyi mutual information
\be
I_n(r_1 : r_2, \mathcal{A}_{r_1 r_2}) \eq S_n(\rho, \mathcal{A}_{r_1}) + S_n(\rho, \mathcal{A}_{r_2}) - S_n(\rho, \mathcal{A}_{r_1 r_2}).
\ee
We find the following phases:
\begin{equation}
    I_n(r_1 : r_2, \mathcal{A}_{r_1 r_2}) \approx \begin{cases}
        0 & n < n_c, \\
         \log \left( q^2 \right) + \frac{n}{1-n} \log  \left(\frac{(p+p')^2}{p''}\right) & n_c < n < n_*, \\
        \frac{n}{1-n} \log \left( \frac{(p' + p'')^2}{p''}\right) & n_* < n.
    \end{cases}
\end{equation}
Now we rewrite the boundary R\'enyi mutual information \er{eq:bdyRenMI} as
\be
S_n(R_1: R_2) = \underbrace{I_n(r_1: r_2, \mathcal{A}_{r_1 r_2}) - I_n(r_1 : r_2, \widetilde{\mathcal{A}}_{r_1 r_2} )}_{\textrm{area contribution}} + \underbrace{I_n(r_1 : r_2, \widetilde{\mathcal{A}}_{r_1 r_2} )}_{\textrm{bulk matter contribution}},
\ee
where the contribution of the nontrivial area operator to the boundary R\'enyi mutual information is identified with the difference of the bulk R\'enyi mutual information in the two algebras.
As stated previously, $I_n(r_1: r_2, \widetilde{\mathcal{A}}_{r_1 r_2})$ is suppressed relative to $I_n(r_1: r_2, \mathcal{A}_{r_1 r_2})$, so this model implements phase transitions in the boundary R\'enyi mutual information without a large bulk matter contribution (in the gauge-invariant algebra). We plot these two phase transitions for an example in Figure \ref{fig:mutual}.

This is a proof of concept showing that adding bulk gauge symmetries to the RTN in this manner allows the boundary R\'enyi mutual information to be nontrivial and $n$-dependent, even for states with small bulk R\'enyi mutual information (in the gauge-invariant algebra).
In our simple example here, the minimal surface does not shift---i.e. it is the same for all $n$---but there is no obstruction to writing a more complicated example in which the location of the minimal surface changes with $n$ due to the nontrivial area operator.

\section{Discussion and Outlook}

In this work, we have presented a modification of the random tensor network which allows us to reproduce known features of semiclassical holographic states. We discuss some open questions and possible future directions below.

We have presented a toy model which, for simple choices of bulk input state, exhibits sharp phase transitions in the R\'enyi entropy and R\'enyi mutual information.  With a sufficiently tuned set of probabilities and irrep dimensions, one could engineer a smooth varying R\'enyi entropy that matches with, for example, the correct one-interval CFT$_2$ R\'enyi entropy \eqref{eq:CFT}. It would be an even more complicated task to reproduce the correct R\'enyi entropy for multiple intervals in the CFT \cite{Hartman:2013mia, Faulkner:2013yia}.

The bulk algebras that we encountered in our model are type I von Neumann algebras. This is in contrast to the type II von Neumann algebras for gravity constructed using the crossed product \cite{Witten:2021unn, Chandrasekaran:2022eqq, Jensen:2023yxy}. A ``type I approximation'' to the crossed product was recently studied in \cite{Soni:2023fke}. It is thus tempting to incorporate the crossed product and the resultant birth of a type II algebra into the tensor network toy models of holography.

Our gauge-invariant subregion algebras generally have nontrivial centers. On the other hand, a prescription was given in \cite{Casini:2013rba} to construct gauge-invariant subregion algebras with trivial centers in lattice gauge theory. This prescription involves adding operators to the algebra that we do not include, so it does not contradict our results in any way.

Here we have implemented a graph version of the lattice gauge theory construction along the lines of Kogut and Susskind, but crucially without dynamics, due to the lack of a Hamiltonian. Because of this, our construction does not have anything more to say about time evolution in tensor networks than previous models. It would be interesting to understand how to incorporate a Hamiltonian and the associated time evolution into tensor networks. It would also be interesting to study the commutators of intersecting area operators in our gauged RTN, which in standard AdS/CFT do not commute \cite{Kaplan:2022orm}.

\section*{Acknowledgements}

We thank Chris Akers, Horacio Casini, David Grabovsky, Daniel Harlow, Kristan Jensen, Don Marolf, and Pratik Rath for interesting discussions. This material is based upon work supported by the Air Force Office of Scientific Research under Award Number FA9550-19-1-0360. This material is also based upon work supported by the U.S. Department of Energy, Office of Science, Office of High Energy Physics, under Award Number DE-SC0011702. SAM would like to thank the Centro de Ciencias de Benasque Pedro Pascal for their hospitality while a portion of this work was completed.

\bibliographystyle{JHEP}
\bibliography{RTN}

\end{document}